\documentclass[11pt,a4paper]{article}
\usepackage{epsfig,graphicx,t1enc,amsmath,subfigure}
\usepackage{amsfonts}
\usepackage[mathscr]{eucal}

\newcommand{\half}{\frac{1}{2}}
\newcommand{\be}{\begin{equation}}
\newcommand{\ee}{\end{equation}}
\newcommand{\ben}{\begin{equation*}}
\newcommand{\een}{\end{equation*}}
\newcommand{\bea}{\begin{align}}
\newcommand{\eal}{\end{align}}
\newcommand{\bean}{\begin{align*}}
\newcommand{\eean}{\end{align*}}
\newcommand{\bes}{\begin{split}}
\newcommand{\ees}{\end{split}}

\newcommand{\idp}{\frac{1}{2}(1+\sigma_3)}
\newcommand{\idm}{\frac{1}{2}(1-\sigma_3)}

\newcommand{\igam}{I\gamma}
\newcommand{\gam}{\gamma}
\newcommand{\isig}{I\sigma}
\newcommand{\sig}{\sigma}

\let\tilde=\widetilde

\begin{document}

\title{A representation of twistors within geometric (Clifford) algebra}

\author{Elsa Arcaute\footnote{Electronic mail: e.arcaute@mrao.cam.ac.uk} \footnote{Current address: Dept. Mathematics, Imperial College London, South Kensington Campus, London, SW7 2AZ, UK.}, Anthony Lasenby\footnote{Electronic mail: a.n.lasenby@mrao.cam.ac.uk} and Chris Doran\footnote{Electronic mail: c.doran@mrao.cam.ac.uk}\\ \emph{Astrophysics group, Cavendish Laboratory,}\\ \emph{JJ Thomson Av., Cambridge, CB3 0HE, UK.}}

\date{}

\maketitle

\begin{abstract}

In previous work by two of the present authors, twistors were re-interpreted as 4-d spinors with a position dependence within the formalism of geometric (Clifford) algebra. Here we extend that approach and justify the nature of the position dependence. We deduce the spinor representation of the restricted conformal group in geometric algebra, and use it to show that the position dependence is the result of the action of the translation operator in the conformal space on the 4-d spinor. 
We obtain the geometrical description of twistors through the conformal geometric algebra, and derive the Robinson congruence. This verifies our formalism.
Furthermore, we show that this novel approach brings considerable simplifications to the twistor formalism, and new advantages.  We map the twistor to the 6-d conformal space, and derive the simplest geometrical description of the twistor as an observable of a relativistic quantum system. The new 6-d twistor takes the r\^ole of the state for that system. In our new interpretation of twistors as 4-d spinors, we therefore only need to apply the machinery already known from quantum mechanics in the geometric algebra formalism, in order to recover the physical and geometrical properties of 1-valence twistors.

\end{abstract}

\section{Introduction}

Twistors are nowadays a versatile mathematical tool, which can be applied to many different areas. Recently, there has been a renaissance in the application of twistors to other fields than the well-known integrable systems. 
In string theory for example, new methods have been developed to compute scattering amplitudes in Yang-Mills theories making use of the twistor space \cite{Berkovits-Witten'04,Witten'04}. These papers (in particular \cite{Witten'04}) have had a very important impact in high energy physics and supergravity theories, leading to extensive research on these areas, see for example \cite{Berkovits'04,C-S-W'04,C-S-W'04_2,C-S-W'04_3,R-V'04,Bed_etal'05,Bidder_etal'05,Bena_etal'05,Bern_etal'05}. 

The initial construction of twistors was motivated by Penrose \cite{Penrose_alg,pen69,pen72,pen77,pen87,pen99,TP} to solve one of the most important and still unresolved problems within theoretical physics: quantum gravity. The twistor formalism relies on a complex structure of the physical world, where massless free fields of general spin are taken as holomorphic functions, and where the geometry of space-time can be related to the principle of superposition in quantum mechanics. 
The building blocks for this last relation are the following crucial aspects of the theory and the conformal space. The twistor algebra is an extension of the spinor algebra, and the twistor space is considered as the most fundamental entity. The spaces needed for general relativity and quantum mechanics, are both subspaces of this more general space. For example, events in space-time are derived in a non-local way through the incidence of twistors.

The main objective of the present paper is to re-interpret twistors and to look at some of the aspects mentioned above, making use of a special case of Clifford algebras: \emph{geometric algebra}. This is a real algebra, that allows for objects such as tensors and spinors, to be represented in a frame-free way.

Clifford algebras have been used to express spinors and twistors for many years. These for example, have been realised as one-sided ideals of Clifford algebras by Cartan, Riesz \cite{riesz}, Chevalley, Atiyah, Bott and Shapiro \cite{ati64}, and Penrose. 
Specific approaches to twistors using real Clifford algebras, can also be found in the literature, see for example \cite{abl82}.
Here however, we follow a different approach. 
This work is based on a previous paper \cite{2-spinors...}, where the authors defined twistors as 4-d spinors with a position dependence, and obtained some of the physical properties of the particle that a 1-valence twistor is encoding. We extend here this formalism, and consider only 1-valence twistors, leaving 2-valence twistors needed for the construction of events in space-time to be treated in \cite{twistors2}. Let us briefly summarise the new results contained in this paper. 

In \cite{2-spinors...}, the nature of the position dependence given to the 4-d spinor was unknown. Here we show that such a position dependence is the outcome of the action of a translation operator on the spinor in the conformal space. 
We construct a representation of the twistor in the 6-d conformal space, and deduce the spinor representation of the restricted conformal group, through the action of the group on the 6-d twistor. This result completes and amends the spinor representation of the restricted conformal group within geometric algebra, previously established in \cite{book} and \cite{proc}. We verify the consistency of our approach, by recovering the main geometrical structure that defines twistors: the \emph{Robinson congruence}. Furthermore, we are able to derive the simplest geometrical representation for a twistor as the observable of a relativistic quantum system, where the 6-d twistor plays the r\^ole of the wave-function for such a system. 

The research described here represents a novel approach to twistors that is in clear contrast with the applications mentioned in the first paragraph, where  holomorphicity plays a fundamental r\^ole, and twistors are used for specific computational tasks only. Furthermore, the use of geometric algebra brings new advantages by simplifying the formalism.

This work is organised as follows. Section 1 is the present introduction. Section 2 contains a general review of the main important results found in \cite{2-spinors...}, and it is divided in 3 subsections. In subsection 2.1 we define and explain the basic general rules of geometric algebra, and specify the algebra for the space-time. In subsection 2.2 we give an outline of spinors following the construction introduced by Hestenes \cite{hes67}. Spinors correspond to objects that belong to the even subalgebra of the Clifford algebra for the space, and that transform in a specific way under the action of the spin group.
Although the even subalgebra has the same dimension as the ideal of the algebra, this will act as a first illustration of how our formalism differs in interpretation and manipulations with the conventional approaches. 
In subsection 2.3 we introduce the definition of a 1-valence twistor, and give the physical properties of the massless particle that it represents. Although the position dependence is justified later on, this last aspect shows that the new re-interpretation simplifies greatly the formalism. In the next section (3), we look at the conformal geometric algebra. This is divided in two parts. The first subsection is a review of how the conformal space is constructed, and how the conformal transformations are defined within geometric algebra. In the second part, we give a novel representation of the twistor in the 6-d space, and use it to deduce the spinor representation of the conformal transformations within geometric algebra. This is achieved by looking at the way the 6-d twistor transforms under the action of the conformal group defined in 3.1. 
This section is crucial, because it shows that the position dependence is simply the result of translating the origin in the conformal space to a general position vector. Furthermore, the geometrical properties of the twistors that we derive in section 4, are constructed through the conformal geometric algebra. 
In 4.1 we describe null twistors geometrically, and recover the expected null ray defined in \cite{PRII}. In 4.2 we look at the geometrical description of non-null twistors, and derive the Robinson congruence. We confirm that the circles belonging to the congruence are geodesics, by proving that these correspond to d-lines in a non-Euclidean space. 
In the final subsection (4.3), we show that by re-interpreting a twistor as  a spinor in geometric algebra, we are able to project it to the 6-d space and use it to obtain its geometrical description, for the simplest case, as one obtains the spin bivector in a relativistic quantum system.\\

The conventions in this work are as follows. Greek indices run from 0 to 3, while Latin indices run from 1 to 3. The signature of the space is chosen such that the timelike vector has positive norm, and the space-like vectors negative. The speed of light is set to unity $c=1$ and so is the Planck constant $\hbar=1$.

\section{Background Review}

\subsection{Space-time algebra}

In this section we introduce our main mathematical tool, which follows Hestenes  approach to Clifford algebras, where no specific representation is specified (see \cite{hes-sta,hes-gc,bigH} for example).
Details of the content of this section can be found in \cite{book}.

\emph{Geometric algebra} (GA) is a Clifford algebra over the field of real numbers. Its bilinear product is given by the \emph{geometric product} defined as follows
\be
ab=a\cdot b+ a\wedge b,
\ee
where $a$ and $b$ are objects of grade 1 (vectors), $a\cdot b$ has grade 0 (a scalar), and $a\wedge b$ returns an object of grade 2 (a bivector). The first operation in the geometric product corresponds to the \emph{inner product}: 
\be
a\cdot b=\half(ab+ba),
\ee
and the second one to the \emph{outer product}:
\be
a\wedge b=\half(ab-ba).
\ee
Elements of the geometric algebra are in general a linear combination of objects of different grade called \emph{multivectors}. The operator defined by: $\langle \ \rangle_k$, projects the object of grade $k$ from the multivector. This allows us to express the general multivector $A$ as follows
\be
A=\langle A \rangle+\langle A\rangle_1+ \ldots=\sum_k \langle A \rangle_k,
\ee
where $\langle \ \rangle=\langle \ \rangle_0$ corresponds to the scalar projection.

The normalised element of highest grade of the algebra is called the \emph{pseudoscalar} and it is denoted by $I$. The reason for this is that it squares to $-1$ in most of the physical spaces of interest. One such a space is Minkowski space-time, which can be generated through the \emph{space-time algebra}. This algebra is spanned by four orthonormal vectors $\{\gam_\mu \}$ which have as matrix representation the Dirac matrices, and therefore obey the Dirac algebra accordingly
\be
\gam_\mu\cdot\gam_\nu=\half(\gam_\mu\gam_\nu+\gam_\nu\gam_\mu)=\eta_{\mu\nu}=\text{diag}(+---), \label{Dirac_alg}
\ee 
where $\mu$ and $\nu$ run from 0 to 3.

The reciprocal vectors are
\be
\gam^0=\gam_0, \qquad  \gam^k=-\gam_k, \qquad k=1,\dots,3.
\ee
The basis elements of the algebra are constructed through the geometric product, and these constitute in total $2^4=16$ elements, viz
\be
{1\atop{\textrm{1 scalar }}} {\{\gamma_{\mu}\}\atop{\textrm{ 4 vectors }}} {\{\gamma_{\mu} \wedge \gamma_{\nu}\}\atop{\textrm{ 6 bivectors }}} {\{I\gamma_{\mu}\}\atop{\textrm{ 4 trivectors }}} {I\atop{\textrm{ 1 pseudoscalar }}}
\ee
where the pseudoscalar $I=\gamma_0\gamma_1\gamma_2\gamma_3$ representing the directed volume element, is such that $I^2=-1$.

For non-relativistic physics, a basis for the 3-d space can be created from the above vectors, by defining \emph{relative vectors}. These are bivectors given by $r\wedge v$, where $r$ is the position vector, and $v$ is the velocity of the observer. Taking $v=\gam_0$ (since $c=1$), the bivectors have the following form
\be
\sigma_k=\gamma_k\gamma_0, \qquad k=1,\dots,3. \label{splitSTA}
\ee
Note that the pseudoscalar is the same for the algebra of the 3-d and 4-d spaces. In a 3-d space the bivectors $\{\sigma_k\}$ can be treated as vectors that obey the Pauli algebra
\be
\sigma_i\sigma_j=\sigma_i\cdot \sigma_j+\sigma_i\wedge \sigma_j=\delta_{ij} +I\epsilon_{ijk} \sigma_k, \label{Pauli_alg}
\ee
where $\delta_{ij}$ is the Kronecker delta, and $\epsilon_{ijk}$ is the Levi-Civita symbol (permutation symbol).
This relation can be verified using eq.(\ref{splitSTA}) and (\ref{Dirac_alg}).

Even elements $R$ of the algebra such that $R\tilde{R}=1$ are called \emph{rotors}. (Note that the tilde over $R$ corresponds to the \emph{reverse} operation, which consists in reversing the order of the elements in each outer product of a multivector.) Any multivector $A$ of the algebra is transformed in exactly the same way under their action
\be
A \ \mapsto \ A'=RA\tilde{R}. \label{RAR}
\ee
However, rotors and spinors $\psi$ transform single-sidedly according to their $4\pi$ symmetry, viz
\be
\psi \ \mapsto \ \psi'=R\psi.\label{rotpsi}
\ee
Rotors mainly encode rotations, and for spaces of Lorentzian signature $(1,q)$ or $(p,1)$, these can be generalise to any dimension since the rotor can be written as follows
\be
R(\lambda)=\pm e^{-\lambda B/2}, \label{R_exp}
\ee
where $\lambda$ denotes the infinitesimal parameter controlling the angle of rotation, and $B$ denotes the plane where the rotation is taking place. More complex transformations can also be defined by rotors. An example of these are the proper orthochronous Lorentz transformations, which can be obtained through the following rotor
\be
R=e^{\alpha \hat{B}/2}e^{\beta I\hat{B}/2},
\ee
where the first exponential generates the boost and the second one the rotation. The bivector $\hat{B}$ encodes the boost, and the rotation takes place in the plane perpendicular to the motion: $I\hat{B}$. The parameter $\beta$ is the difference of velocities between the frames (in units of $c$), and $\alpha$ is related to it as follows
\be
\tanh\alpha=\beta, \qquad (\beta<1).
\ee

Rotors form a subgroup of the spin group, which consists of even-grade multivectors $S \in \mathcal{G}(p,q)$ such that
\be
SaS^{-1} \in \mathcal{G}^1, \quad \forall a\in \mathcal{G}^1, \quad S\tilde{S}=\pm 1,
\ee
where $\mathcal{G}(p,q)$ denotes the geometric algebra for a space with signature $(p,q)$, and $\mathcal{G}^1$ stands for the subspace of elements of grade 1 of the algebra. The spin group $Spin(p,q)$ is a 2-to-1 map of the group of orthogonal transformations with unitary determinant $SO(p,q)$. Taking into account that $R^{-1}=\tilde{R}$, rotors constitute the spin subgroup such that $R\tilde{R}=1$, and this is designated by $Spin^+(p,q)$. It is a double-cover representation of the restricted orthogonal group $SO^+(p,q)$, which is the subgroup of transformations continuously connected with the identity.

\subsection{Spinors}

Within geometric algebra, spinors belong to the even subalgebra and to the spin group (see \cite{hes-sta,hes90,STA&e,book}). In the 3-d space they are called Pauli spinors, and they are proportional to rotors. The relation between the conventional notation and our definition is
\be
|\zeta \rangle ={\zeta^0 \choose \zeta^1}={a^0+ia^3 \choose -a^2+ia^1} \quad \leftrightarrow \quad \zeta=a^0+a^k I\sigma_k, \qquad k=1,\dots,3,  \label{Pauli_spinor}
\ee
where the coefficients $a^\mu$ (for $\mu=0,\ldots,3$) are scalars.
Note that the Einstein summation convention is applied when two indices repeat themselves, unless otherwise stated.
Using the above equivalence, we can write the corresponding spin-up and spin-down basis states 
\be
|\uparrow \rangle \quad \leftrightarrow \quad 1 , \qquad \quad |\downarrow \rangle \quad \leftrightarrow \quad -I\sigma_2. \label{spin-up/down}
\ee
In terms of these, the spinor $\zeta$ in eq.(\ref{Pauli_spinor}) can be expressed as
\be 
\zeta=(a^0+a^3I\sigma_3)+(-I\sigma_2)(-a^2+a^1I\sigma_3), \label{psi_updown}
\ee
which indicates that the r\^ole of the unit imaginary is taken by the bivector $\isig_3$. Therefore, borrowing the same notation to express the components in terms of GA, according to eq.(\ref{psi_updown}) these correspond to
\be 
\begin{array}{ll}
\zeta^0=a^0+a^3\isig_3\\[2mm] 
\zeta^1=-a^2+a^1 \isig_3
\end{array} 
\ee
Every time we encounter $\isig_3$ in an expression for spinor components, we just replace it by the unit imaginary $i$ in order to obtain the conventional notation. 
In general, if a given spinor $\zeta$ is expressed in its geometric algebra representation, its components are recovered through the following operation
\be
\begin{array}{ll}
\zeta^0=\langle \zeta \rangle + \langle -\zeta \ \isig_3 \rangle \isig_3=\langle \zeta \rangle_{0,\isig_3}=\langle \zeta \rangle_s \\[2mm] 
\zeta^1=\langle \isig_2 \ \zeta \rangle + \langle -\isig_2 \ \zeta \ \isig_3 \rangle \isig_3=\langle \isig_2 \ \zeta \rangle_s
\end{array} 
\ee
where the subindex $s$ in the projector operator indicates that the scalar and $I\sigma_3$ terms are projected from the spinor $\zeta$. The operator $\langle \ \rangle_s$ corresponds to the projector used to construct the inner product.

Note that $\zeta \tilde{\zeta}=\rho$, where $\rho$ is a scalar. This enables us to write the spinor in terms of a rotor $R$ as follows: $\zeta=\rho^{1/2} R$, since $R\tilde{R}=1$. This form justifies why spinors behave as operators within our formalism, as we will see later on.

Looking now at the relativistic scenario, we are interested in two particular spinors: 4-d spinors and 2-spinors (or Weyl spinors). The former have the same algebraic structure as Dirac spinors, and hence have 8 degrees of freedom, while the latter although relativistic, have only 4 degrees of freedom. Dirac spinors and 4-d spinors are two different entities, even though they have the same algebraic structure, because they do not transform in the same way under the conformal group as we will see. However, they share other important properties that will allow us to define quantum mechanical observables in the same way for both of them. 4-d spinors can be expressed in terms of the 2-spinors in the same way that Dirac spinors can, and this is called the Weyl representation. In the literature, different approaches can be found for representations of 2-spinors in terms of Clifford algebras (see for example \cite{FK}). Here we follow the approach first introduced in \cite{2-spinors...}, and later on in \cite{2,Chris,book}. Weyl spinors are irreducible representations of $SL(2,\mathbb{C})$ (see \cite{Buchbinder} for example). The fundamental representation denoted by $(\half,0)$, corresponds to the left-handed Weyl spinor: $|\omega\rangle=\omega^A$, where $A=0,1$. In terms of GA this is
\be
\omega^A \quad \leftrightarrow \quad \omega \frac{1}{2}(1+\sigma_3), \label{2-spinor}
\ee
where $\omega$ is a Pauli spinor, i.e. of the form of eq.(\ref{Pauli_spinor}). The projector to the right of $\omega$ is a chiral operator that ensures that the spinor maintains its four degrees of freedom after undergoing Lorentz transformations.

The complex conjugate representation is given by $(0,\half)$, and it corresponds to the right-handed representation which lives in a different module. This spinor transforms in a different way to the left-handed one. We choose to define the operation of complex conjugation by right multiplication by $\sig_1$, since this changes the ideal which defines the module. The right-handed spinor $|\overline{\omega}\rangle=\bar{\omega}^{A'}$ is therefore given in terms of GA by
\be
\bar{\omega}^{A'} \quad \leftrightarrow \quad \omega \frac{1}{2}(1+\sigma_3)\sigma_1 = \omega I\sigma_2 \frac{1}{2}(1-\sigma_3). \label{omega_cc}
\ee

To obtain a representation of the 2-spinor in terms of components, it is necessary to define a \emph{spin-frame}. This is given by a pair of spin-vectors $\underline{o}$ and $\underline{\iota}$, which are normalised
\be
\{\underline{o},\underline{\iota}\}=1.
\ee
The operation $\{ \ , \ \}$ denotes the inner product which is defined further on.
Explicitly, the spin-vectors are $\underline{o}=(1,0)$ and $\underline{\iota}=(0,1)$. In terms of GA, the map of eq.(\ref{spin-up/down}) tells us that the corresponding Pauli spinors of the spin-frame are: $o=1$ and $\iota=-I\sigma_2$, which define the spin-frame for 2-spinors as follows
\bea
o^A & \quad \leftrightarrow  \quad \idp, \\
\iota^A & \quad \leftrightarrow  \quad -I\sigma_2\idp.
\end{align}
The expression for a general spin-vector in terms of its components is 
\be
\underline{\omega}=(\omega^0,\omega^1) \quad \text{or} \quad \omega^A=\omega^0o^A+\omega^1\iota^A. \ee
These are recovered from the GA representation as follows
\be
\begin{array}{ll}
\omega^0=2 \langle \omega \ \idp \rangle_s \\[2mm] 
\omega^1=2 \langle \isig_2 \ \omega \ \idp \rangle_s
\end{array}  \label{2-spinors_cpts}
\ee
while the components of its complex conjugate $\bar{\omega}^{A'}$, defined by eq.(\ref{omega_cc}), are
\be
\begin{array}{ll}
\bar{\omega}^{0'}=2 \langle  \omega \ \idm \rangle^*_s \\[2mm] 
\bar{\omega}^{1'}=2 \langle \isig_2 \  \omega \ \idm \rangle^*_s
\end{array} \label{2-spinors_cpts_cc}
\ee
where the projector $\langle \ \rangle^*_s=\langle \ \rangle_{0,-\isig_3}$ denotes the complex conjugate.

The inner product between 2-spinors is an antisymmetric quantity, which takes the following form in terms of components 
\be
\omega_A\pi^A=\omega_0\pi^0+\omega_1\pi^1=\omega^0\pi^1-\omega^1\pi^0,
\ee
where the relation between representations of components 
\be
\omega^0=\omega_1, \qquad \omega^1=-\omega_0, \label{w^0w_1}
\ee
was used. Note that this relation is always valid for any 2-spinor $\omega^A$.

The geometric algebra representation of the inner product corresponds to
\be
\omega_A\pi^A=\alpha+i\beta \quad \leftrightarrow \quad {\{\omega,\pi}\}= \alpha+\beta \isig_3 =\langle I\sigma_2\tilde{\omega}\pi\rangle_s \label{GAinner_spinors},
\ee
and the complex conjugate is simply denoted by ${\{\omega,\pi}\}^*$.

We proceed to define the Weyl representation of the 4-d spinor, viz
\be
|\psi\rangle={|\omega\rangle \choose |\overline{\pi}\rangle} \quad \leftrightarrow \quad \psi=\omega\frac{1}{2}(1+\sigma_3)+\pi I\sigma_2 \frac{1}{2}(1-\sigma_3).  \label{Dspinor}
\ee
Its algebraic similarity with the Dirac spinor, enables us to recover the bilinear covariants, in the same way that we recover those for the Dirac spinor, by applying the machinery of quantum mechanics in the geometric algebra formalism. We can therefore define the `Dirac adjoint' of this 4-d spinor, and this is 
\be
\langle\bar{\psi}| \quad \leftrightarrow \quad \tilde{\psi},
\ee
which denotes a fully Lorentz-covariant operation. The components of $\psi$ can be recovered using the following projections
\be
\begin{array}{ll}
\psi^0=\omega^0=2 \langle \psi \ \idp \rangle_s \\[2mm]
\psi^1=\omega^1=2\langle \isig_2 \ \psi \ \idp \rangle_s \\[2mm]
\psi^2=\bar{\pi}_{0'}=-\bar{\pi}^{1'}=-2 \langle \psi \ \idm \rangle_s\\[2mm]
\psi^3=\bar{\pi}_{1'}=\bar{\pi}^{0'}=-2\langle \isig_2 \ \psi \ \idm \rangle_s
\end{array} \label{Weyl_cpts}
\ee
It is important to bear in mind that every time the bivector $\isig_3$ appears in an expression for components, we need to replace it by $i$, the unit imaginary, in order to recover the conventional notation.

The spinors take the r\^ole of operators within the geometric algebra formalism. This is elucidated when the observables of a system are found.  One of the fundamental aspects to recover such observables in the relativistic framework, is the action of the Dirac matrices on the spinor. In our matrix-free representation, such an action is equivalent to
\bea
\hat{\gamma}_{\mu}|\psi\rangle & \quad \leftrightarrow  \quad \gamma_{\mu}\psi\gamma_0 ,\qquad (\mu=0,\ldots,3) \label{gam_psi}\\ 
i|\psi\rangle & \quad \leftrightarrow  \quad \psi I\sigma_3 , \label{i_psi} \\ \hat{\gamma}_5|\psi\rangle & \quad \leftrightarrow  \quad \psi\sigma_3,  \label{gam_5}
\end{align}
where $\psi$ is a Dirac or a 4-d spinor.

Another fundamental aspect is the inner product. For Pauli, Dirac and 4-d spinors this is given by
\be
(\psi,\phi)_s =\langle \tilde{\psi} \phi\rangle_s=\langle\tilde{\psi}\phi\rangle-\langle\tilde{\psi}\phi I\sigma_3\rangle I\sigma_3. \label{()_s}
\ee
In the space-time, this operation leads to a space with signature $(2,2)$, and with a structure equivalent to a complex space. This is a crucial point for future identification of twistors with 4-d spinors in our formalism, since these are exactly the properties of the twistor space. 

Let us now illustrate how the spinors correspond to operators in GA by recovering some observables. Even though we are considering 4-d spinors denoted by $\psi$, the bilinear covariants that we will find below will be denoted and called in the same way as their analogues in the Dirac theory, since both objects have exactly the same properties. The Dirac current for example, is obtained as follows
\be
\mathcal{J}_{\mu}=\langle\bar{\psi}|\hat{\gamma}_{\mu}|\psi\rangle \quad \leftrightarrow \quad 
(\psi,\gam_\mu \psi \gam_0)_s= \langle \tilde{\psi} \gam_\mu \psi \gam_0 \rangle =\gamma_{\mu}\cdot \mathcal{J},\label{J}
\ee
where $\mathcal{J}=\psi\gamma_0\tilde{\psi}$ denotes the Dirac current in the geometric algebra formalism. In terms of components, vectors are represented as follows within GA
\be
\mathcal{J}=\mathcal{J}^\mu\gam_\mu=\mathcal{J}_\mu\gam^\mu, \label{cpts_vector}
\ee
which tells us that $\mathcal{J}_\mu=\mathcal{J}\cdot\gam_\mu$ or $\mathcal{J}^\mu=\mathcal{J}\cdot\gam^\mu$.
Equation (\ref{J}) indicates that $\psi$ acts as a rotor, and gives the instruction to rotate $\gamma_0$ in the direction of the current, and dilate it. 

Another quantity of interest is the relativistic generalisation of the spin vector, which is the relativistic angular momentum. This is called the spin bivector in GA, and we find it as follows
\be
S^{\mu\nu}=\half \langle  \bar{\psi}|i\half[\hat{\gam}^\mu,\hat{\gam}^\nu]| \psi\rangle \quad \leftrightarrow \quad -\half (\psi,\gam^\mu\wedge\gam^\nu \ \psi\isig_3)_s=-S\cdot (\gam^\mu\wedge\gam^\nu)
\ee
where
\be
S=\half  \psi I\sigma_3\tilde{\psi}, \label{spin_bivector}
\ee
is the spin bivector. Since this object is in no specific representation, in order to identify it with its conventional analogue, we need to express it in terms of components.
From the above equation we see that we chose to define the components for second rank tensors as follows
\be 
S^{\mu\nu}=-S\cdot (\gam^\mu\wedge\gam^\nu), \label{S^mn}
\ee
although any other choice could have been made.\\

The twistor algebra, which is an extension of the spinor algebra, can be taken as the fundamental structure for quantum mechanics and general relativity \cite{PRII}, because the tensor algebra can be derived from the spinor algebra.
For example, the group property that $SL(2,\mathbb{C})$ is the universal covering group of the restricted Lorentz group $SO^+(3,1)$, enables us to represent null vectors through the outer product of left- and right-handed Weyl spinors. In detail (see \cite{PRI}), the spinor system is built up from two modules $\mathfrak{G}^{A}$ and $\mathfrak{G}^{A'}$ which are related to each other by a relation of complex conjugation. The tensor algebra can be constructed using the outer product of combinations of the different modules. Here we are interested in defining the \emph{flagpole}, which is a future pointing null vector $K^{AA'}$, that represents the 2-spinor up to a phase. This is a fundamental object to describe the twistor physically and geometrically. On one hand, the geometrical information of a twistor can be specified through the flagpole directions of the primary part of the twistor. And on the other, the flagpole of the projected part gives the momentum of the massless particle that the twistor is encoding. Within GA, the flagpole of a spinor $\omega^A$ is defined as follows (see \cite{book} and \cite{Chris})
\be
K^{AA'}=\omega^A\bar{\omega}^{A'} \quad \leftrightarrow \quad K=\frac{1}{2}\omega(\gamma_0+\gamma_3)\tilde{\omega}. \label{flagpole}
\ee
This has an interesting form within this formalism, since it corresponds to the Dirac current $\mathcal{J}=\phi\gamma_0\tilde{\phi}=K$ associated to the wave-function $\phi=\omega\frac{1}{2}(1+\sigma_3)$. 
The components of $K$ as a vector can be recovered through eq.(\ref{cpts_vector}). However, this vector can also be represented in terms of $2\times 2$ Hermitian matrices, which is the conventional representation of $K^{AA'}$. The representation we choose here is
\be
K^{AA'}=K^\alpha\sig_\alpha=\left(\begin{array}{cc} K^{00'} & K^{01'} \\[2mm] K^{10'} & K^{11'} \end{array} \right)=\left(\begin{array}{cc} K^0+K^3 & K^1-iK^2 \\[2mm] K^1+iK^2 & K^0-K^3 \end{array} \right), \label{K^AA'}
\ee
where $\sig_0=\mathbf{1}$ and $\sig_k$ are the Pauli matrices as usual.
This representation is the same used by \cite{Buchbinder}, although it differs from the one specified in \cite{PRI}, so special care has to be taken when components are compared. (The representation we use can be obtained following the method developed in \cite{PRI}, although the stereographic projection has to be taken this time from the South Pole.)

\subsection{Twistors and their physical properties}\label{twist_phys}

The re-interpretation of a twistor as a 4-d spinor in the geometric algebra formalism can take place, because the position dependence that a twistor has and a spinor normally does not, is introduced here through the action of an operator on the 4-d spinor. Such an operator corresponds to a translation in the conformal space, as we will see later on.

Twistors (see \cite{PRII}) are objects belonging to a 4-d complex space, which can be described through the solutions to the equation
\be
\nabla_{A'}^{(A}\omega^{B)}=0.
\ee
For non-charged fields in Minkowski space $\mathbb{M}$, the solution is
\begin{equation} 
\begin{array}{ll} \omega^A=\stackrel{\scriptscriptstyle{o}}{\omega}^A-ir^{AA'}\stackrel{\scriptscriptstyle{o}}{\pi}_{A'} \\[2mm] 
\pi_{A'}=\stackrel{\scriptscriptstyle{o}}{\pi}_{A'} \end{array}  \label{sol_twist}
\end{equation}
where $\stackrel{\scriptscriptstyle{o}}{\omega}^A$ and $\stackrel{\scriptscriptstyle{o}}{\pi}_{A'}$ are constant spinor-fields whose values coincide with those of $\omega^A$ and $\pi_{A'}$ at the origin, and $r^{AA'}$ is a vector field on $\mathbb{M}$. This solution is encoded in a \emph{twistor} as follows
\be
Z^{\alpha}=(\omega^A,\pi_{A'}).
\ee
In terms of GA, the twistor is represented through the following construction
\be
Z=T_{-r}(\psi)=\psi+r\psi \igam_3 \idp, \label{twistor}
\ee
where $\psi$ is the 4-d spinor given by eq.(\ref{Dspinor}), and the operator $T_{-r}$ is the spinor representation in the conformal space of a translation to $-r$. This will be elucidated later on.

The 2-spinor $\omega^A$ with the locational properties is called the \emph{primary part} of the twistor, and the constant 2-spinor $\pi_{A'}$ its \emph{projection} part. Each of these parts plays an important r\^ole in the geometrical and physical interpretation of twistors as we will shortly see. 
In terms of GA, the primary part corresponds to
\be
\omega^A \quad \leftrightarrow \quad \omega_P=(\omega+r\pi \isig_2 \igam_3)\idp 
=Z\idp \label{primary}
\ee
where $\omega \idp$ is just a constant spinor-field, and we can verify that its value coincides with that of $\omega_P$ at the origin as established. Its components can be recovered making use of eq.(\ref{2-spinors_cpts}), however note that $r^{AA'}$ must be expressed according to eq.(\ref{K^AA'}). 
On the other hand, the projection part is
\be
\pi_{A'} \quad \leftrightarrow \quad Z\idm=\pi \isig_2\idm.
\ee
For comparison of this GA object with the conventional formalism, we need to resort to components. Eq.(\ref{2-spinors_cpts_cc}) gives us a way to recover those for $\pi^{A'}$, and using eq.(\ref{w^0w_1}) we can obtain those for $\pi_{A'}$.

The components for the twistor are conventionally evaluated at the origin, see eq.(6.1.21) on p.48 of \cite{PRII}. Therefore, within the GA framework these correspond to eq.(\ref{Weyl_cpts}), viz
\be
Z^0=\psi^0, \quad Z^1=\psi^1, \quad Z^2=Z_{0'}=\psi^2, \quad Z^3=Z_{1'}=\psi^3.
\ee 

Taking a twistor as a `translated' 4-d spinor, enables us to find its physical properties in the same way we find the observables of a quantum system described by $\psi$. To do this we need to define the inner product. This is the same as for 4-d spinors given by eq.(\ref{()_s}), since the conformal operation defined by $T_r$ preserves it as we will see later on. This is therefore
\be
X^{\alpha}\bar{Z}_{\alpha} \quad \leftrightarrow \quad 
\langle \tilde{X}Z\rangle_s=\langle \tilde{\phi}\psi\rangle_s,
\ee
where
\be
X=\phi +r\phi I\gamma_3\frac{1}{2}(1+\sigma_3) , 
\ee
and 
\be
\phi=\xi\idp+\eta I\sigma_2\idm.
\ee
The observables obtained through this operation are therefore conformally invariant. One such a quantity is the \emph{helicity}, which is defined as
\be
\bes 
s&= -\langle \tilde{Z}Z \rangle_s \\
&= -\langle \tilde{\psi}\psi \rangle.\\ 
\end{split} \label{s}
\ee
This is fully independent of any point in space-time and asserts a handedness that divides the twistor space into three regions according to $s>0$, $s<0$, or $s=0$.

All the physical states in quantum field theory can be labelled according to the eigenvalues of two Casimir operators: the momentum and the Pauli-Lubanski spin vector. The linear momentum is a future-null vector field, which in terms of 2-spinors represents the flagpole direction of the projection part of the twistor. According to eq.(\ref{flagpole}) this is
\begin{align} \nonumber
p_{AA'}&=\bar{\pi}_A\pi_{A'} \\  \nonumber &\updownarrow \\  p&=\half \pi(\gamma_0+\gamma_3)\tilde{\pi}= \half \psi(\gamma_0-\gamma_3)\tilde{\psi}=\half Z(\gamma_0-\gamma_3)\tilde{Z} \label{p^a}
\end{align}
which is a covariant quantity as well.

The angular momentum has a position dependence given by a conformal Killing vector field, which for the massless case corresponds to the flagpole directions of $\omega^A$. We will soon see that these directions have an important geometrical interpretation. They correspond to the tangents to the lines of the congruence that defines a twistor up to a scalar factor. In terms of GA, the angular momentum takes the same form as in the quantum theory 
\be
M=\half ZI\sigma_3\tilde{Z}, \label{M_Z}
\ee
which in terms of the position and the momentum is
\be
M=M_0-r\wedge p, \label{M}
\ee
where
\be
M_0=\half \psi I\sigma_3\tilde{\psi} \label{M0}
\ee  
is the angular momentum at the origin as expected, since it corresponds to the spin bivector of the spinor $\psi$, defined by eq.(\ref{spin_bivector}). This last equation confirms that $M$ given by eq.(\ref{M_Z}) is of the form of a quantum observable. Its components can be recovered making use of eq.(\ref{S^mn}).

The Pauli-Lubanski spin vector is defined as follows within GA
\be
S=-2 I(p\wedge M)=2 p.(IM).
\ee
For massless particles this can be expressed as
\be
S=-p\langle \tilde{\psi}\psi \rangle=ps,
\ee
where $s$ is the helicity of the particle.

Now that we have seen how to obtain essential physical quantities from twistors, we can reverse our procedure. A specific twistor $Z$ can be reconstructed up to a phase if we know $p$ and $M$. This is easy to see by replacing $Z$ with $Ze^{I\sigma_3\theta}$ (where $\theta$ is a scalar), and verifying that the quantities remain unchanged. \\

In this section we have shown how to recover within our framework the physical properties of the massless particle that the twistor is encoding, if this is re-interpreted as a 4-d spinor. The formalism led to expressions already known from relativistic quantum mechanics. 
In the next section we will define the conformal space and derive the spinor representation of the conformal transformations, justifying the form of the twistor as a `translated' 4-d spinor. Furthermore, the conformal space is crucial for the geometrical interpretation of twistors.

\section{Conformal geometric algebra}

\subsection{Conformal space and its transformations}

The conformal space consists in the addition of two new directions, $e$ and $\bar{e}$, perpendicular to the basis vectors of the original space $V(p,q)$,
\be
e^2=1, \qquad \bar{e}^2=-1, \qquad e\cdot \bar{e}=e\cdot x=\bar{e}\cdot x=0,  \label{e}
\ee
where $x \in V(p,q)$. Therefore, for a vector space $V$ of signature $(p,q)$, we get a vector space $V(p+1,q+1)$, where two null directions $n$ and $\bar{n}$ are formed from the additional vectors
\be
n=e+\bar{e}, \qquad \bar{n}=e-\bar{e}. \label{nnb}
\ee
Note that,
\be
n\cdot \bar{n}=2 \quad \textrm{and} \quad x\cdot n=x\cdot \bar n=0. 
\ee

Each position vector $x \in V(p,q)$ is as a consequence represented in the conformal space by a null vector $X \in V(p+1,q+1)$, through a map that is specific to the geometry of the space. 

The construction of the Euclidean conformal map within geometric algebra was first given by \cite{hes-gc}, and this is defined as follows (see \cite{book})
\be
X=F_E\left(\frac{x}{\lambda}\right)=\frac{1}{2\lambda^2}(x^2n+2\lambda x-\lambda^2\bar{n}), \label{F_E(x)}
\ee
where $\lambda$ is a positive scalar, a fundamental length scale, introduced in order to obtain a consistent dimensionless object. 

For the hyperbolic space, the map is given by \cite{CG_univ}
\be
X=F_H\left(\frac{x}{\lambda}\right)=\frac{1}{\lambda^2-x^2}(x^2n+2\lambda x-\lambda^2\bar{n}). \label{F_H(x)}
\ee

The properties of the space are determined through the \emph{conformal transformations}, which are transformations that preserve angles. These define a group $C(p,q)$, which is a double-cover representation of $SO(p+1,q+1)$, and has its same dimension. There is a subgroup of transformations that can be defined in terms of infinitesimal parameters. These specify the restricted group $C^+(p,q)$, where only inversions are excluded. The transformations of $C^+(p,q)$ can be expressed in terms of rotors, since the group of rotors is a double-cover representation of the restricted orthogonal group. 

The results presented here are taken from \cite{book}, modified to include the scale factor $\lambda$ in order to have the correct dimensionality.

\begin{description}

\item {\bf Translations}

The operation we want is
\be
x\mapsto x'=x+a
\ee
where $x,a\in V(p,q)$. This is achieved in the conformal space with the rotor
\be
T_a=\exp\left(\frac{na}{2\lambda}\right)=1+\frac{1}{2\lambda} na, \label{T_a}
\ee
since
\be
T_a \ F_E\left(\frac{x}{\lambda}\right) \ \tilde{T}_a=F_E\left(\frac{x+a}{\lambda}\right). \label{translations}
\ee
Note that this rotor leaves the point at infinity, given in the Euclidean space by $n$, invariant: $T_an\tilde{T}_a=n$.

In the hyperbolic space, the translation rotor is given by (see \cite{CG_univ})
\be
T_x=\frac{1}{\sqrt{\lambda^2-x^2}}(\lambda+\bar{e}x). \label{T_aH}
\ee
The form of this rotor indicates that as $x$ approaches the value of $\lambda$, the point is translated to infinity. This implies that $\lambda$ is setting the \emph{boundary} of the representation, which in 2-d corresponds to the circle of circumference of the disc. This representation is known as the \emph{Poincar\'e disc}, and it is a convention to take a unit disc: $\lambda=1$. 

An important property of a curved space, is that general translations do not commute: $T_xT_y\neq T_yT_x$, unless $x$ and $y$ are parallel.

\item {\bf Rotations}

Rotors $R$ in charge of rotations belong to the space-time algebra. As a consequence, the point in the conformal space is rotated in the same way as the multivectors of the space-time algebra. Therefore, according to eq.(\ref{RAR}), the transformation is the following
\be
F_E\left(R\frac{x}{\lambda}\tilde{R}\right)=RF_E\left(\frac{x}{\lambda}\right)\tilde{R}.
\ee
Note that the point at infinity is left invariant under the action of the rotor.

If instead of a rotation about the origin, we want a rotation about the point $a\in V(p,q)$, the operation is achieved with the new rotor
\be
R_a=T_aR\tilde{T}_a,
\ee
which is equivalent to translating $a$ back to the origin, performing the rotation, and then translating it forward again. The covariance of the theory is therefore manifested, since the origin is not a special point.

\item  {\bf Dilations}

A dilation in the origin is given by
\be
x \mapsto x'=e^{-\alpha}x, 
\ee
where $\alpha$ is a scalar. The rotor encoding such a transformation in the conformal space is
\be
D_{\alpha}=e^{\frac{\alpha N}{2}}=\cosh\left(\frac{\alpha}{2}\right)+\sinh\left(\frac{\alpha}{2}\right)N, \label{D_alpha}
\ee
where $N=e\bar{e}$. The transformation in the conformal space is
\be
e^{-\alpha}D_{\alpha}F_E\left(\frac{x}{\lambda}\right)\tilde{D}_{\alpha}=F_E\left(e^{-\alpha}\frac{x}{\lambda}\right). \label{dilations}
\ee

If we want a dilation about a point $a$, we proceed as we did for the rotations, and obtain a new rotor, viz
\be
D_{\alpha}'=T_aD_{\alpha}\tilde{T}_a=e^{-\frac{\alpha A\wedge n}{2}},
\ee
where $A=F_E\left(\frac{a}{\lambda}\right)$.

\item {\bf Inversions}

This operation cannot be defined infinitesimally, and therefore, cannot be expressed in terms of a rotor.

Taking into account the scale factor, the operation we need to consider is
\be
\frac{x}{\lambda} \ \mapsto \ \frac{\lambda x}{x^2}.
\ee
Under inversions the point at the origin and at infinity are exchanged. This can be achieved through reflections, where the vector $e$ is taken to be the unit vector to the plane. This operation leads to
\be
-eF_E\left(\frac{x}{\lambda}\right)e=\frac{x^2}{\lambda^2}F_E\left(\frac{\lambda}{x}\right). \label{inversions}
\ee
Note that this is the correct operation, since vectors in the conformal space are represented homogeneously, therefore $X$ and $\alpha X$ correspond to the same vector $x$, for any non-null scalar $\alpha$.

Furthermore, inversions in an arbitrary point $a$ are obtained by replacing the vector $e$ by $T_ae\tilde{T}_a$.

\item {\bf Special conformal transformations}

This operation is defined as an inversion followed by a translation, and another inversion. The position vector is transformed to
\be
\bes
\frac{x}{\lambda} \quad \stackrel{\text{inv.}}{\longrightarrow} \quad x_i=\frac{\lambda x}{x^2} \quad \stackrel{\text{trans.}}{\longrightarrow} \quad x_{it}&=\frac{\lambda x}{x^2}+\frac{a}{\lambda} \\ 
& \qquad \quad \big\downarrow \text{\scriptsize{inv.}} \\ 
x'&=\left(\frac{\lambda^2 x+ax^2}{\lambda x^2}\right)^{-1} \\
&=x\frac{\lambda}{\lambda^2+ax} =\frac{\lambda}{\lambda^2+xa}x 
\end{split}
\ee
The generator of this transformation is also a rotor
\be
K_a=eT_ae=1-\frac{\bar{n}a}{2\lambda}, \label{K_a}
\ee
which leads to the following conformal vector
\be
K_aF_E\left(\frac{x}{\lambda}\right)\tilde{K}_a=\left(1+\frac{2}{\lambda^2}a\cdot x+\frac{1}{\lambda^4}a^2x^2\right)
F_E\left(x\frac{\lambda}{\lambda^2+ax}\right). \label{SCT}
\ee

\end{description}

In conclusion, the operators achieving restricted conformal transformations in GA belong to the conformal rotor group $Spin^+(2,4)$. This group is a double-cover representation of the restricted orthogonal group $SO^+(2,4)$, which in its turn is a double-cover representation of the restricted conformal group $C^+(1,3)$, see \cite{book} p.384. Therefore, our rotor group is a 4-fold covering of the latter. This is a key point, because $Spin^+(2,4)$ is isomorphic to $SU(2,2)$, the twistor group.

\subsection{Twistors in the 6-d conformal space}

Twistors are a spin-1/2 representation of the restricted conformal group. We can therefore map the twistors into the 6-d conformal space-time algebra, apply the rotors to the 6-d new objects and deduce the spinor representation for the conformal group. Furthermore, we will see in the next section, that the geometrical description of the simplest twistor can be constructed as an observable of this special 6-d spinor in the conformal space.

The twistor is mapped into a 6-d object as follows
\be
\Upsilon=Z W_1W_2,
\ee
where $W_1$ and $W_2$ are the following projector operators
\be
W_1=\frac{1}{2}(1-I\gamma_3 e), \qquad W_2=\frac{1}{2}(1-I\gamma_0 \bar{e}).
\ee

Let us now look at the induced transformations on the twistor $Z$, by transforming $\Upsilon$ under the conformal group. We take $\lambda=1$.

\begin{description}

\item {\bf Rotations}

These are straightforward, since the rotor is defined in the 4-d space, and therefore the transformation verifies
\be
\bes
R\Upsilon&=R(ZW_1W_2)\\
&=(RZ)W_1W_2.\\
\end{split}
\ee
In this case the rotor acts in the usual way and does not take a different form
\be
Z \ \mapsto \ R_0(Z)=RZ, \label{spin_rot}
\ee
where $R_0$ denotes a rotation in the origin. To obtain rotations about an arbitrary point, we need to use translations.

\item {\bf Translations}

Let us apply the translation rotor defined by eq.(\ref{T_a}) to the 6-d twistor
\be
\bes
T_a\Upsilon&=T_a(ZW_1W_2)\\
&=ZW_1W_2-\frac{a}{2}Z(e+\bar{e})W_1W_2.\\
\end{split}
\ee
If we compute the action of the vectors $e$ and $\bar{e}$ on the projectors, we find that
\be
eW_1W_2=I\gamma_3W_1W_2 \quad \text{and} \quad \bar{e}W_1W_2=-I\gamma_0W_1W_2,
\ee
and therefore
\be
\bes
T_a\Upsilon&=ZW_1W_2- a Z I\gamma_3\idp W_1W_2 \\
&=\left(Z-a Z I\gamma_3\idp\right) W_1W_2.\\
\end{split}
\ee
This tells us that the spinor representation for translations
is
\be
T_a(Z)=Z-aZ I \gamma_3 \idp, \label{spin_trans}
\ee
since
\be
T_a\Upsilon=T_a(Z)W_1W_2.
\ee
Eq.(\ref{spin_trans}) therefore confirms that the twistor $Z$ corresponds to a translated 4-d spinor to $-r$ within the geometric algebra framework.

Note that this operation leaves the inner product invariant, as expected. If we set
\be
\psi'=T_a(\psi) \qquad \text{and} \qquad \phi'=T_a(\phi)
\ee
it follows that
\be
(\psi',\phi')_s=(\psi,\phi)_s ,\label{()_sinv}
\ee
where $\psi$ and $\phi$ are general 4-d spinors.

\item  {\bf Dilations}

The action of the rotor for dilations defined by eq.(\ref{D_alpha}) on $\Upsilon$ is
\be
\bes
D_{\alpha}\Upsilon&=D_{\alpha}(ZW_1W_2)\\
&=\cosh\left(\frac{\alpha}{2}\right)ZW_1W_2+\sinh\left(\frac{\alpha}{2}\right)Z(-\sigma_3)W_1W_2\\
&=Z e^{-\alpha \sigma_3/2}W_1W_2.\\
\end{split}
\ee
This tells us that general 4-d spinors are transformed under dilations as follows
\be
D_{\alpha}(Z)=Z e^{-\alpha\sigma_3/2}, \label{spin_dil}
\ee
since
\be
D_{\alpha}\Upsilon = (D_{\alpha}Z) W_1W_2.
\ee
Note that the above operation represents a dilation in the origin. If we want dilations about a general point, we need to apply translations as well.

\item {\bf Inversions}

This transformation is not part of the restricted conformal group, and therefore it is not given in terms of a rotor operator. Furthermore it does not have a unique representation. For this case, we look directly at the spinor representation first. We have a freedom of choices for the operator $\hat{O}$ defining the inversion, for example combinations of $I\sigma_2$ and $I\sigma_1$.
The operator $\hat{O}$ needs to be \emph{anti-unitary} and such that
\be
D_{\alpha}(Z \hat{O})=D_{-\alpha}(Z)\hat{O}. \label{inv_cond}
\ee
We choose $I\sigma_2$, which leads to
\be
Z \mapsto Z'=Z I\sigma_2, \label{spin_inv}
\ee
and this agrees with eq.(\ref{inv_cond}) since $I\sigma_2$ anticommutes with $\sigma_3$.

Let us now find the appropriate transformation of $\Upsilon$ that would lead to $ZW_1W_2 \mapsto ZI\sigma_2 W_1W_2$ according to eq.(\ref{spin_inv}). The choice is not unique since the inverse transformation of $Z$ is neither unique. This is a behaviour expected from $\Upsilon$, since it is a spinor. We choose the following operation
\be
\bes
-e\Upsilon I\gamma_1 &= -e Z W_1W_2 I\gamma_1 \\
&=Z I \sigma_2 W_1W_2\\
\end{split}
\ee
which clearly works.

\item {\bf Special conformal transformations}

This transformation is achieved through the rotor given by eq.(\ref{K_a}). Its action on the 6-d twistor leads to
\be
\bes
K_a \Upsilon &= K_a (ZW_1W_2)\\
&= ZW_1W_2+ a Z I\gamma_3\idm W_1W_2 \\
&=\left(Z+ a Z I\gamma_3\idm\right) W_1W_2.\\
\end{split}
\ee
Spinors therefore transform under special conformal transformations as follows
\be
K_a(Z)=Z+aZ I \gamma_3 \idm, \label{spin_SCT}
\ee
since
\be
K_a \Upsilon  =  (K_aZ)W_1W_2.
\ee
Note that due to the single-sided operation of the rotor, we get a flaw in the sign of the transformation coming from the anti-unitarity of inversions, viz 
\be
\bes
\Upsilon \quad \stackrel{\text{inv.}}{\longrightarrow} \quad -e\Upsilon I\gamma_1 \quad \stackrel{\text{trans.}}{\longrightarrow} \quad &\left(1+\frac{1}{2}(e+\bar{e})a\right)(-e\Upsilon I\gamma_1) \\ & =-e\left(1-\frac{1}{2}\bar{n}a\right) \Upsilon I\gamma_1\\
& \qquad \qquad \ \ \big\downarrow \text{\footnotesize{inv.}}\\
& \left(1-\frac{1}{2}\bar{n}a\right) \Upsilon (-1)  =  -K_a\Upsilon.
\end{split}
\ee
This shows that we cannot have a faithful representation of the restricted conformal group in terms of spinors.

Such a flaw naturally also takes place for 4-d spinors
\be
\bes
Z \quad \stackrel{\text{inv.}}{\longrightarrow} \quad Z I\sigma_2 \quad \stackrel{\text{trans.}}{\longrightarrow} & \quad  Z I\sigma_2-aZ I\sigma_2 I\gamma_3 \idp  \\
& \qquad \qquad \qquad \ \ \big\downarrow \text{\footnotesize{inv.}}\\
& -Z-aZ I\gamma_3 \idm  =  -K_a(Z).
\end{split}
\ee
This flaw in the sign is therefore indicating that these spinors are a 4-valued representation of the restricted conformal group. This agrees with the definition of a 1-valence twistor as a 4-valued spinor representation of the restricted conformal group.

\end{description}

Spinors are formed from even elements of the space-time algebra, and the rotors of the restricted conformal transformations are constructed from bivectors of the form $\gam_\mu\gam_\nu$ for spatial rotations and boosts, and of the form $e\gamma_{\mu}$ and $\bar{e}\gamma_{\mu}$ for the rest. We have found that we can summarise the above transformations, by stating the representation of the action of these bivectors on a general 4-d spinor $\psi$
\bea
e\gamma_{\mu} & \quad\leftrightarrow \quad -\gamma_{\mu}\psi I\gamma_3 \label{egam}\\
\bar{e}\gamma_{\mu} & \quad\leftrightarrow \quad -I\gamma_{\mu}\psi \gamma_0. \label{ebgam}
\end{align}
These are the accurate maps to use, and this amends the ones previously established in \cite{proc} and \cite{book}, which only differ by a minus sign.

The formalism described in this section is the keystone to the geometric algebra approach to twistors, since not only it justifies the position dependence of the twistor as a 4-d spinor, which comes about just by translating the origin to a general position vector, but also, the conformal geometric algebra allows us to construct the geometrical structure of twistors in a simple way.

\section{Geometrical description of twistors}

In this section we reproduce the results found in \cite{PRII,Penrose_alg} for the geometrical description of twistors, making use of our formalism, where twistors are 4-d spinors. 

The information in a twistor can be encoded geometrically, defining the twistor up to a scalar. We encounter two different types of twistors according to their scalar product, which describes the helicity eq.(\ref{s}). 
If this one vanishes the twistor is called \emph{null} and otherwise \emph{non-null}. 

Furthermore, the twistor can be represented geometrically in two different ways which are equivalent, since they are the dual of each other. One of them consists in obtaining the locus representing the twistor in the complexified (or real) space-time, while its dual is given by the flagpole field of the primary part of the twistor which is called the \emph{Robinson Congruence}.

\subsection{Null twistors: $\langle \tilde{Z} Z \rangle=0$}

The case of a null twistor is the simplest, and its representation can be obtained in the real space-time. This is a null ray, which points in the flagpole direction of $\pi$, and passes through the point $q$ which lies along the flagpole direction of $\omega$. This representation is achieved for the case when the primary part of the twistor is null. Let us see this in detail. Using eq.(\ref{primary}), we get the following equation for the locus
\be
\omega_P=0 \quad \Leftrightarrow \quad \omega \idp+r\pi \isig_2 \igam_3\idp=0. \label{locus}
\ee
We can multiply this equation on the right by $\gam_0\tilde{\omega}$ to obtain
\be
\omega\half(\gam_0+\gam_3)\tilde{\omega}=-r\pi\sig_2\idm\tilde{\omega}.
\ee
The left-hand side of this equation is the flagpole direction of $\omega$, according to eq.(\ref{flagpole}), and we will denote it by $K$. A particular solution $q$ can be found if we evaluate the equation at $s=0$, which is the condition for null twistors, and this is
\be
q=\frac{K}{\beta}, \label{q}
\ee
where
\be
\beta=-\isig_3 \{\omega,\pi \}^*\Big|_{s=0}. \label{beta}
\ee
The general solution can be expressed as
\be
r=q+hp, \label{ray}
\ee
where $h$ is a real scalar and $p$ is the momentum given by eq.(\ref{p^a}). This can be verified easily, since
\be
p\left(\pi\sig_2\idm\tilde{\omega}\right)=0.
\ee
The solution given by eq.(\ref{ray}) therefore represents the position vector of a null straight line, a \emph{ray}. This is in the direction of the flagpole of $\pi$ and passes through the point $q$, which can be pictured as the position vector of the point of intersection of the line with the null cone at the origin. Note that the ray remains invariant under rescaling: $Z \mapsto Z'=\lambda Z$, where $\lambda$ is just a scalar of the space-time algebra. This indicates that the twistor can only be determined by its ray up to proportionality. 

The case that we just considered, is for special null twistors, such that their primary part is zero. If instead we consider now the case where the projection part vanishes, it is not possible to find a finite locus. For a non-trivial twistor, the locus can be taken as the `light cone at infinity'.

\subsection{Non-null twistors: $\langle \tilde{Z} Z \rangle \ne 0$}

In this case the locus, given by a null line as well, is in the complexified space. A real realization can be obtained through the dual picture, which corresponds to the \emph{system of all null lines which meet the locus} \cite{Penrose_alg}. This system is the \emph{Robinson Congruence}, and is defined by the field of flagpole directions of the primary part $\omega_P$.

The congruence can be visualised by taking a particular example, since any two Robinson congruences can
be transformed one into the other by a Poincar\'e transformation. We will obtain a congruence of rays that \emph{twist} about one another without shear, in a right-handed way if the helicity is positive and in the other sense otherwise. Furthermore, we will confirm the geodetic property of these circles, by showing that these correspond to d-lines in a non-Euclidean space. 

Let us use the same example of \cite{PRII} p.59-63 to see how the congruence is found in terms of GA. The particular twistor denoted here by $Z_{eg}$, corresponds to
\be
Z_{eg}=\psi_{eg}+r\psi_{eg}\igam_3\idp,
\ee
where
\be
\psi_{eg}=-\isig_2 s\idp+\isig_2\idm,
\ee
and
\be
r=t\gamma_0+x\gamma_1+y\gamma_2+z\gamma_3.
\ee
In this example $s$ denotes the helicity. This can be verified using eq.(\ref{s})
\be
-\langle\tilde{Z}_{eg}Z_{eg}\rangle_s=s.
\ee

Let us now find the field of flagpole directions of the primary part of the twistor, since this gives the tangent field to the Robinson congruence. The primary part is
\be
\omega_P=Z_{eg}\idp=(-\isig_2s+r\isig_2\igam_3)\idp,
\ee
and its flagpole defined by eq.(\ref{flagpole}) will be denoted by $K$.
In order to get a 3-d picture of the tangent field, and therefore of the congruence, we need to project $K$ into a hyperplane. 
We choose to do such manipulations in the conformal space making use of the results found in chp.10 of \cite{book}, since the computations are easily achieved there. 
We first obtain the line in the $K$ direction, and then we project it. The line in the conformal space is of the form
\be
L=Ke\bar{e}+r\wedge K \wedge n,  \label{line6d}
\ee
where $r$ is the position vector of a point in the line, $e$ and $\bar{e}$ are the added vectors to construct the conformal space, eq.(\ref{e}), and $n$ is the null vector of eq.(\ref{nnb}). This line is projected into the hyperplane $t=\tau$, which is represented in the conformal space by 
\be
P=(\gamma_0+\tau n)I_6,
\ee
where $I_6=\gam_0\gam_1\gam_2\gam_3 e\bar{e}$ is the pseudoscalar of the 6-d space. The projected line in the plane is
\be
L_P=L+PLP,
\ee
since $PLP$ corresponds to the reflection of the line $L$ with respect to the plane $P$. 

\begin{figure}[t]
\subfigure[View at $45^{\circ}$]{\label{field} 
\begin{minipage}[t]{0.5\linewidth}
\centering
\includegraphics[height=6cm,width=5.5cm]{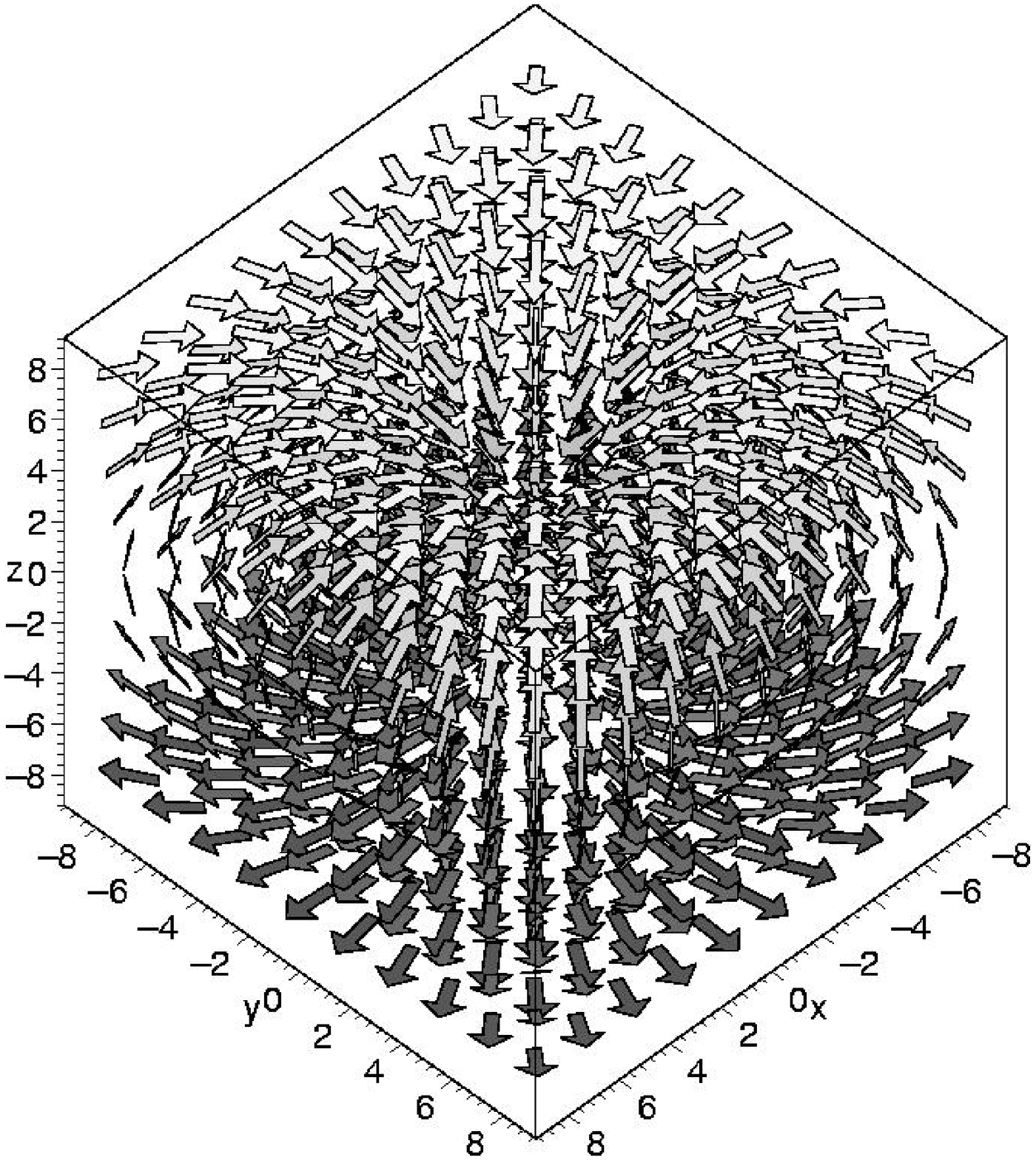}
\end{minipage}}%
\subfigure[Frontal view]{\label{front_field}
\begin{minipage}[t]{0.5\linewidth}
\centering
\includegraphics[height=4.8cm,width=5.3cm]{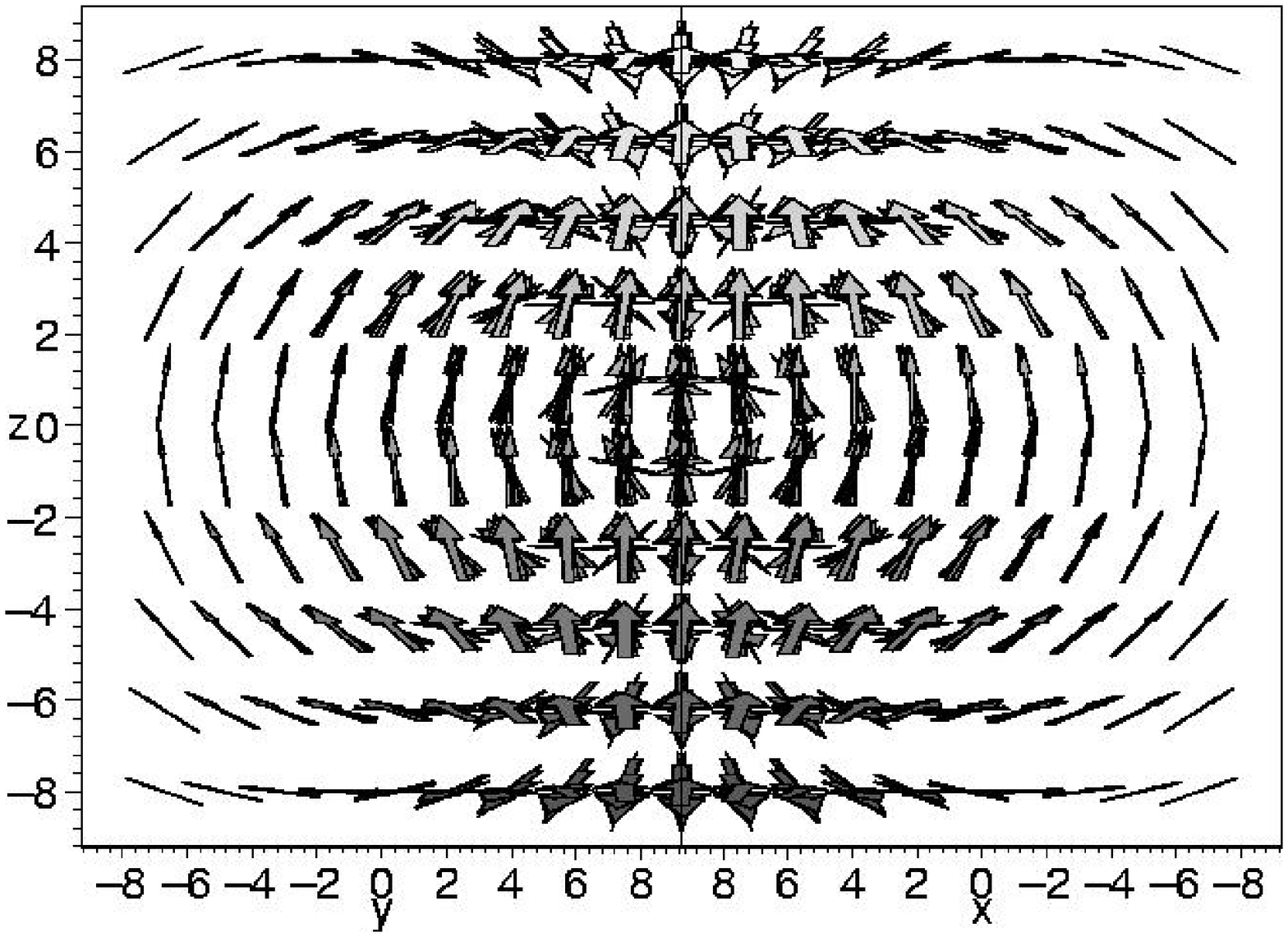}
\end{minipage}}
\caption{ Projected tangent field from two different views \ref{field} and \ref{front_field}.}\label{tangent_field}
\end{figure}

The direction of the line can be recovered easily if the line passes through the origin. This is so because according to eq.(\ref{line6d}) we see that at the origin  we have: $L_o=Ke\bar{e}$. Thereupon, let us translate the line $L_P$ and make it pass through the origin to recover its direction.
The translation rotor that we need here to make the line pass through the origin is (recall eq.(\ref{T_a}))
\be
R_{Bo}=1-\frac{1}{2}nr(t=\tau),
\ee
where $r(t=\tau)$ is just the position vector $r$ evaluated at the intersection $t=\tau$.
The direction of the projected tangent can now be obtained
\be
T_{dir}=(R_{Bo}L_P\tilde{R}_{Bo})e\bar{e}. \label{T_dir}
\ee
To plot this vector field (its normalised version: $T_{ndir}$), we need to set values for $\tau$ and $s$. Let us take for example $\tau=0$ and $s=\frac{1}{2}$. The following plots are obtained fig.\ref{tangent_field}. 

If we compare this tangent field with the figure of the Robinson congruence of \cite{PRII} on p.62, we see that it has the expected form. Furthermore it is important to note that the congruence advances in opposite direction to the $z$ axis, i.e. in the projected direction of the flagpole of $\bar{\pi}^A$, which corresponds to the projected momentum.

\begin{figure}[t]
\subfigure[Positive helicity: $s=10$]{\label{spin+}
\begin{minipage}[t]{0.5\linewidth}
\centering
\includegraphics[height=6cm,width=5.5cm]{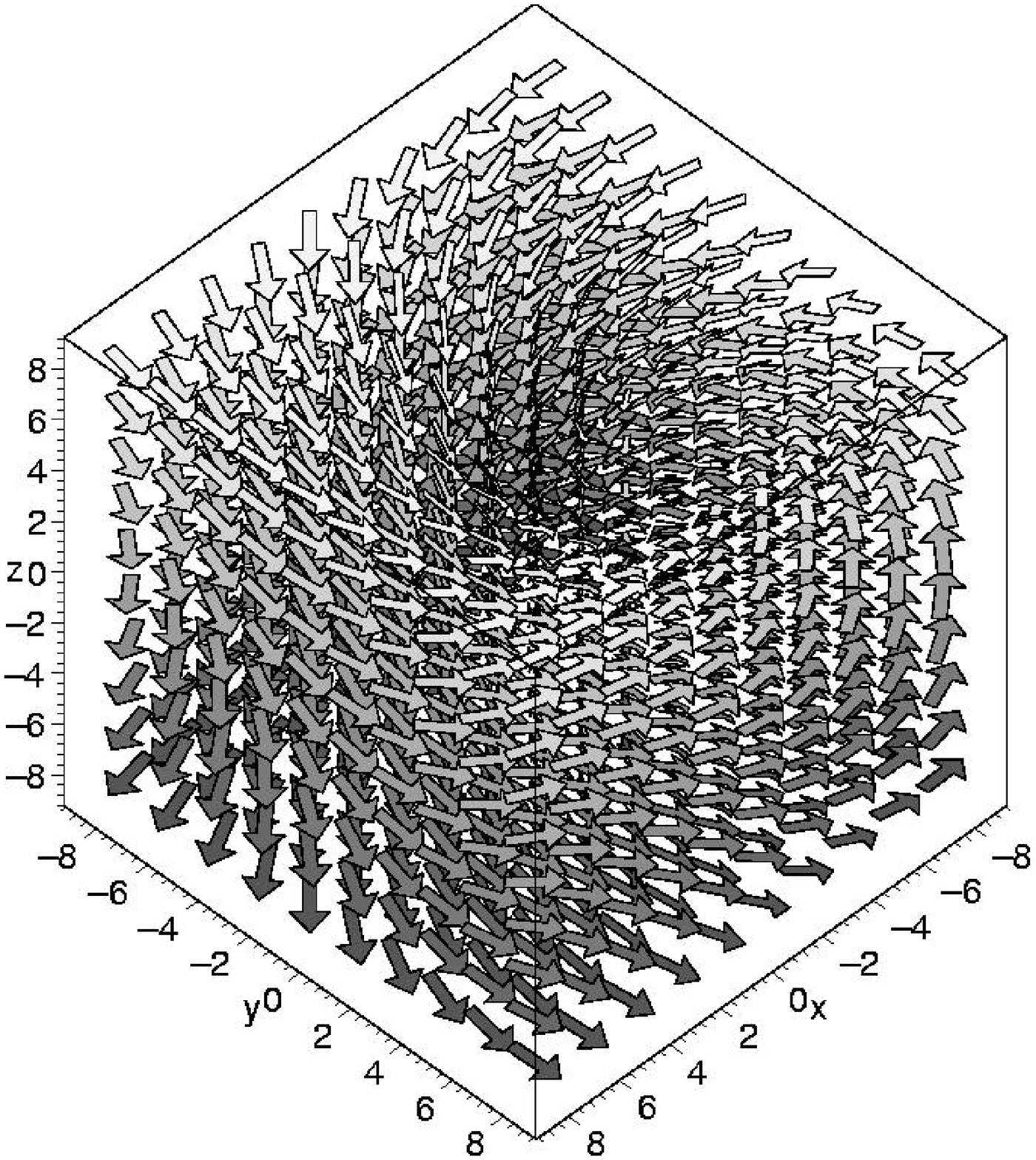}
\end{minipage}}%
\subfigure[Negative helicity: $s=-10$]{\label{spin-}
\begin{minipage}[t]{0.5\linewidth}
\centering
\includegraphics[height=6cm,width=5.5cm]{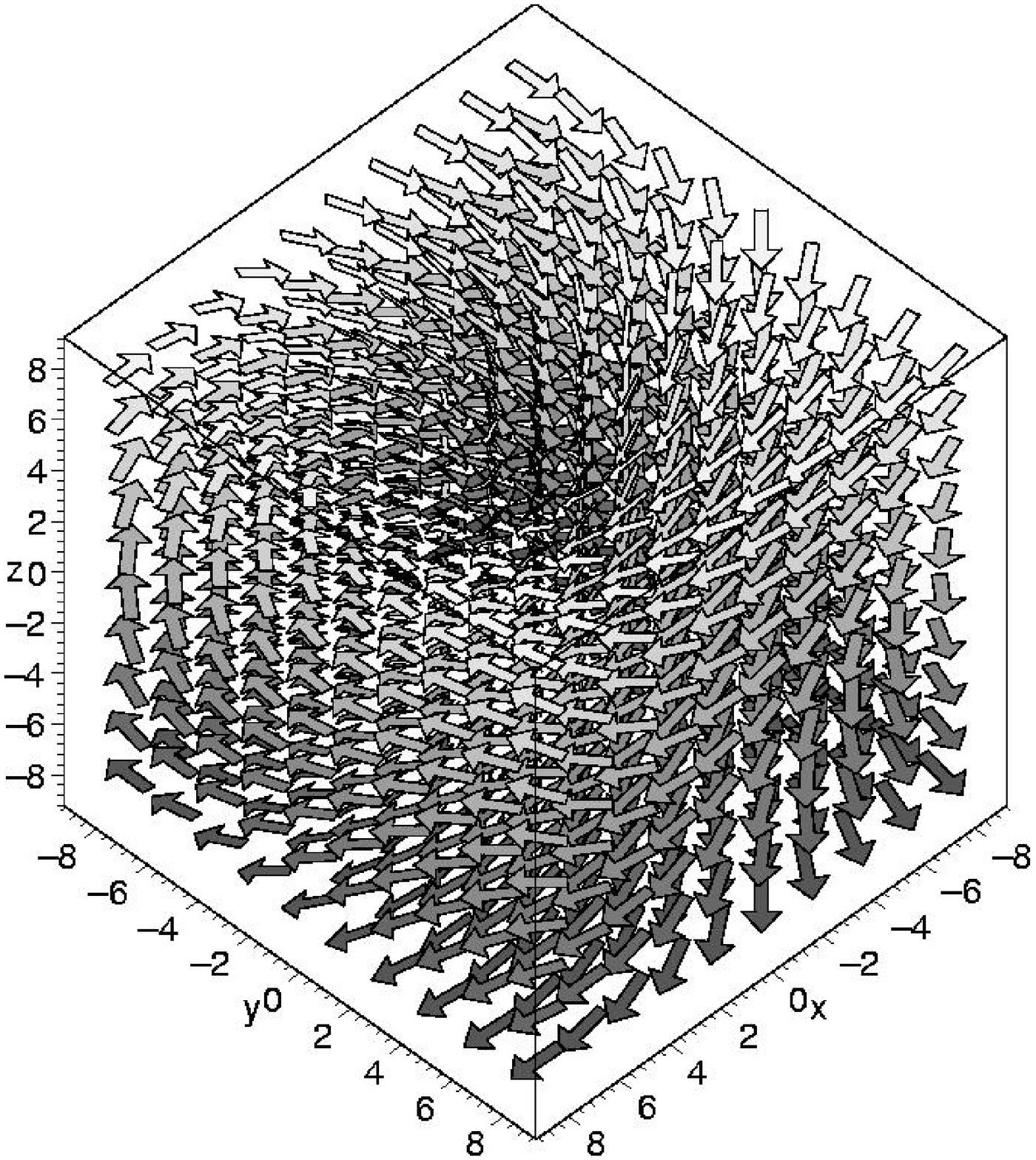} 
\end{minipage}}
\caption{Tangent field at $45^{\circ}$ for positive and negative helicities.}\label{spin+-}
\end{figure}

In order to make evident the r\^ole of the helicity in the twisting of the curves, let us plot two different figures for exaggerated helicities: $10$ and $-10$. Fig.\ref{spin+-} shows the projected tangent field viewed at $45^{\circ}$ for the two cases. We can see that the twisting of the lines is right-handed for positive helicity, fig.\ref{spin+}, and in the opposite direction for negative helicity, fig.\ref{spin-}.

The curves with tangent field $T_{ndir}$ are constructed as follows. First we take the relative vector of $r$, in order to be able to visualise the curves in 3-d space, and parametrise it, viz
\be
r_{rel}(\mu)=r(\mu)\wedge\gamma_0=x(\mu)\sigma_1+y(\mu)\sigma_2+z(\mu)\sigma_3. \label{r_rel}
\ee
The tangent field defines the velocity
\be
\frac{\partial r_{rel}(\mu)}{\partial \mu}=v(\mu)=T_{ndir}(\mu)(t=0)\wedge\gamma_0, \label{xdot}
\ee
and hence the acceleration is given by: $a(\mu)=\frac{\partial v(\mu)}{\partial \mu}$.

After verifying that the motion is confined to a plane, circular and with constant acceleration, we can define the centre of the circle as follows
\be
c_T=r_{rel}(\mu)+|\rho|\frac{a(\mu)}{|a(\mu)|},
\ee
where $|\rho|$ is the magnitude of the radius of the circle, which for this sort of motion is $1/|a(\mu)|$, since the velocity is $|v(\mu)|=1$.

To obtain the circle we need to rotate the radius $\rho=r_{rel}(\mu)-c_T$ from $\theta=0$ to $\theta=2\pi$. The position vector of any point in the circle is therefore
\be
r_{circ}=c_T+R\rho\tilde{R},
\ee
where $R$ is the rotor in the plane
\be
R=\cos\left(\frac{\theta}{2}\right)+B\sin\left(\frac{\theta}{2}\right).
\ee
$B$ is the bivector encoding the plane where the rotation takes place
\be
B=v(\mu)\wedge \frac{a(\mu)}{|a(\mu)|},
\ee
and we see that it is normalised: $B^2=-1$.

\begin{figure}[t]
\subfigure[View at $45^{\circ}$]{\label{45_circles}
\begin{minipage}[t]{0.5\linewidth}
\centering
\includegraphics[height=4.5cm]{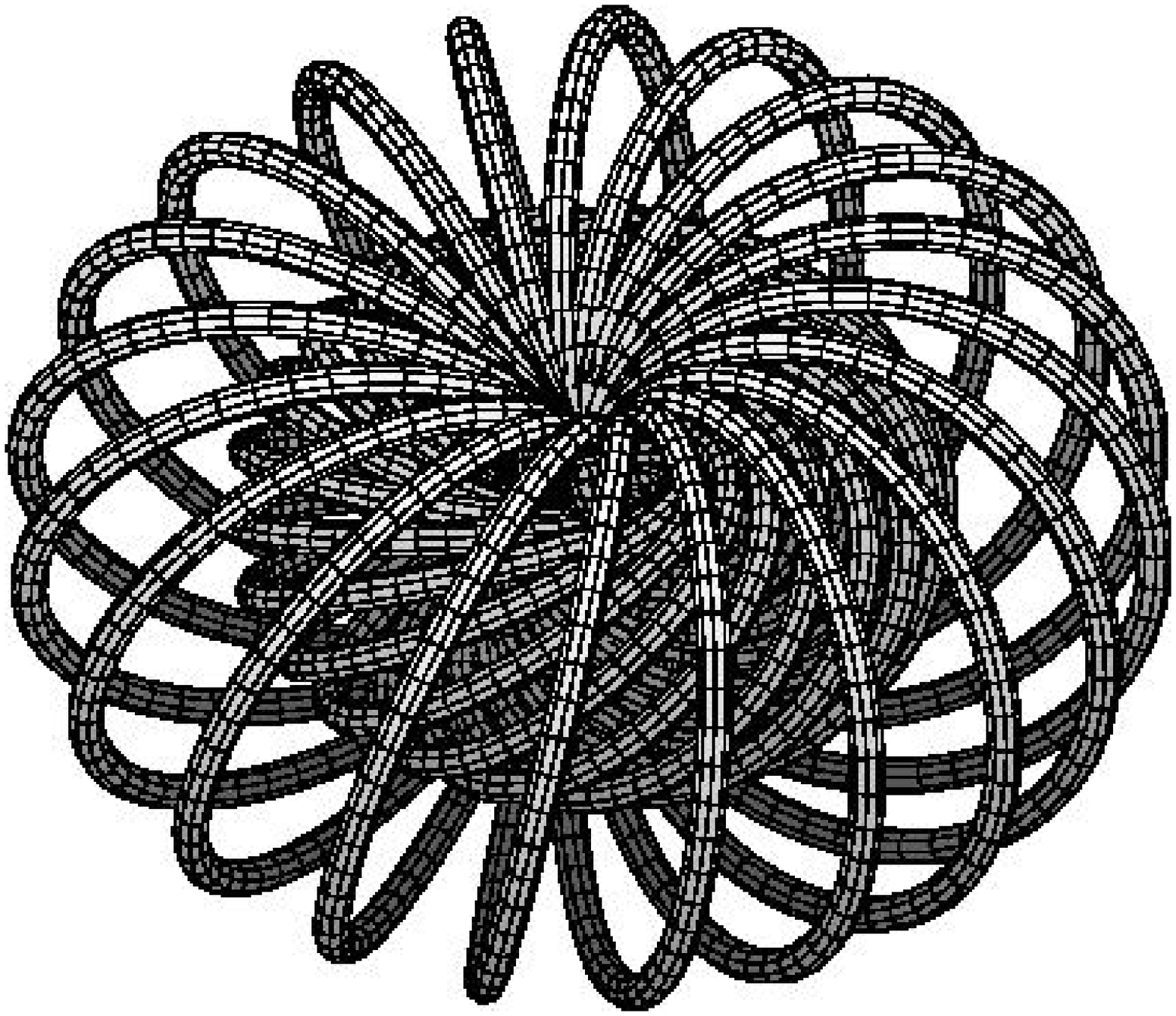} 
\end{minipage}}%
\subfigure[Upper view]{\label{upper}
\begin{minipage}[t]{0.5\linewidth}
\centering
\includegraphics[height=4.5cm]{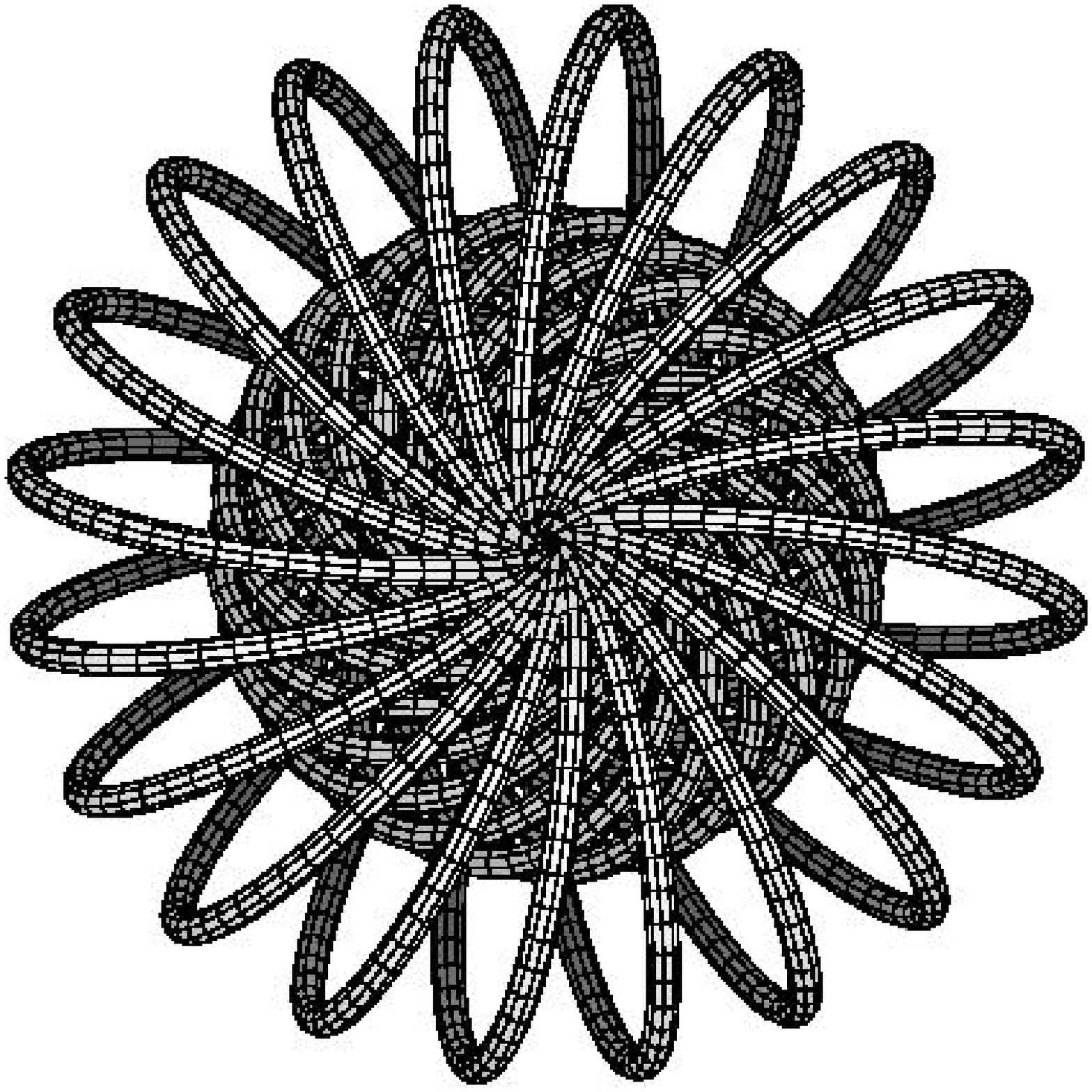}
\end{minipage}}
\caption{Congruences of circles from two different perspectives.}\label{cong}
\end{figure}

To get a plot of the circle we take specific values of $x=N_x$, $y=N_y$, $z
=N_z$ and $s$, with $\theta \in [0,2\pi]$. This circle is part of a family of circles around a torus. To get more members of this family we take
\be
\bes
x&= N_x\cos(\phi), \\
y&= N_y\sin(\phi), \\
z&= N_z,\\
\end{split}
\ee
and vary $\phi$ from an initial angle $\phi_i$ to $\phi_f=\phi_i+2\pi$. 
The following plots are obtained for two different sets of initial coordinates and helicities.
These circles do not intersect at all. Fig.\ref{cong} shows the congruences\footnote{The lines in the plots were chosen as tubular for the sake of perspective, in order to be able to distinguish each family of circles.} from two different perspectives.

The Robinson congruence has therefore successfully been recovered within the formalism of geometric algebra. We now confirm the geodetic property of the curves, by showing that these correspond to d-lines in the non-Euclidean space.

The conformal representation of the position vector $u_{circ}$ in a non-Euclidean space, is given according to eq.(\ref{F_H(x)}) by
\be
X_{circ}=\frac{1}{s^2-u_{circ}^2} (u_{circ}^2n+2su_{circ}-s^2\bar{n})
\ee
where $u_{circ}=r_{circ}\cdot \gamma_0$ for consistency with the non-Euclidean space. Note that the helicity $s$ is identified with the fundamental length scale $\lambda$ introduced to make the null vectors $X$ in the conformal space dimensionless. This is crucial in our formalism, since for cosmological scenarios, this constant is identified with the cosmological constant \cite{CG_univ}! Therefore such a constant tells us about the fundamental structure of the space, and in this case, the helicity is the responsible of the underlying geometry of the twistor.

\begin{figure}[t]
\centerline{\includegraphics[height=6cm,width=4cm]{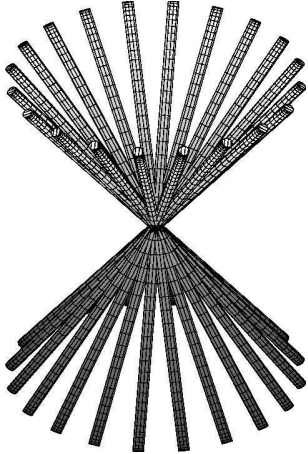}}
\caption{\label{cone} Congruence of d-lines at the origin.}
\end{figure}

Let us now translate the circles to the origin in order to confirm their nature as d-lines. The rotor achieving translations in the conformal space is of the form of eq.(\ref{T_aH}), therefore for this specific task it corresponds to
\be
T_{-u_{circ}}=\frac{1}{\sqrt{s^2-u_{circ}^2}}(s-\bar{e}u_{circ}). \label{T_u}
\ee
The new position vector in the conformal space is thereupon
\be
X'_{circ}=T_{-u_{circ}}X_{circ}\tilde{T}_{-u_{circ}}.
\ee
From this we can recover the 3-d position vector $u'_{circ}$ as follows
\be
u'_{circ}=\sum_{k=1}^3 s \frac{X'_{circ}\cdot\gam_k \ \gam_k}{X'_{circ}\cdot n}.
\ee
If we plot the family of circles obtained from this position vector, proceeding in exactly the same way as before, we find a cone through the origin! See fig.\ref{cone}. 

This result confirms that the circles of the Robinson congruence are of geodetic nature, since the geodesics in a hyperbolic space are represented conformally as circles that become straight lines at the origin (see for example \cite{bigPenrose}, \cite{Brannan} and \cite{anth_geom}).

\subsection{Null ray as an observable of the 6-d space}

Re-interpreting a twistor as a 4-d spinor within geometric algebra, has allowed us to apply the machinery already known from quantum mechanics, to recover the physical properties of the massless particle encoded in a twistor. 
In this section, we want to show that a further advantage of this new interpretation, is that the geometrical properties of a twistor can also be obtained as quantum observables. This takes place in the conformal space, and therefore, the quantum system is defined by the 6-d twistor $\Upsilon$. 

Let us consider the simplest geometrical description, which is the null ray representing a particular null twistor. The equation for the ray is eq.(\ref{ray}), and this can be expressed in the conformal space by making use of eq.(\ref{line6d}), which gives us
\be
L=q\wedge p\wedge n+p e\bar{e}.
\ee
Any object proportional to $L$ will define a line passing through $q$ in the direction $p$.

We set the observer at the origin, which means that $\Upsilon$ is replaced by
\be
\Psi=\psi W_1W_2, \label{6d_spinor}
\ee
where $\psi$ is the 4-d spinor in eq.(\ref{twistor}).
Since the helicity is the object that defines the geometry of the twistor in space-time, let us see if we can find the ray in terms of the spin bivector. According to eq.(\ref{spin_bivector}), this would be of the form
\be
L_{\psi}=(\Psi I\sigma_3\tilde{\Psi})\wedge n.
\ee
We can expand this expression taking into account the results found in \ref{twist_phys} for $\psi$, and we obtain
\be
L_{\psi}=\frac{1}{2}(M_0\wedge n + p e \bar{e}),
\ee
where $M_0$ is the angular momentum at the origin given by eq.(\ref{M0}).

The total angular momentum of a twistor, given by eq.(\ref{M}), is zero when the twistor is null. In this case the angular momentum at the origin $M_0$ can be expressed in terms of the momentum $p$ and the position vector $q$, viz
\be
M_0\bigg|_{s=0}=q\wedge p \bigg |_{s=0}.
\ee
The observable takes now the following form
\be
L_{\psi}=\frac{1}{2}(q\wedge p \wedge n+p e\bar{e}),
\ee
which leads to
\be
L=2L_{\psi},
\ee
if the condition for null twistors is satisfied.

In conclusion, the ray representing a particular null twistor can be obtained as an observable of a quantum system in 6-d, where the state is described by the spinor $\Psi$, and the observable is related to the spin bivector. 

Let us look at interesting features of this, by moving around our observable.

If we apply a translation we find
\be
\bes
L_{\psi}'&=T_a L_{\psi}\tilde{T}_a \\
&=(\Psi' I\sigma_3\tilde{\Psi}')\wedge n \\
&=\half[(q+a)\wedge p \wedge n +p e\bar{e}]\\
\end{split}
\ee
where $\Psi'=T_a \Psi$. This new ray is still in the direction of $p$, but passes now through the point $q+a$.

If we now apply an inversion to the 6-d spinor, let us see how the ray transforms
\be
\bes
L_{\psi}'&=(\Psi' I\sigma_3\tilde{\Psi}')\wedge n \\
&=-e[(\Psi I\sigma_3\tilde{\Psi})\wedge \bar{n}]e \\
&=\half[P\wedge K \wedge n +K e\bar{e}]\\
\end{split}
\ee
where $K$ is the flagpole direction of $\omega$ as before, and $P$ corresponds to the point of intersection of the new null ray with the null cone. This is given by
\be
P=\frac{1}{\beta}p,
\ee
where $\beta$ is defined by eq.(\ref{beta}), and is the same coefficient as the one for the point $q$ given by eq.(\ref{q}).
This indicates that the `inverted' new ray, which is in the direction $K$ passing through $P$, is perpendicular to the old one, and therefore it belongs to the family of rays that are dual to this observable.

\section{Conclusions}

Throughout this work we have shown that twistors can be consistently re-interpret in terms of 4-d spinors with a position dependence, within the framework of geometric algebra. The manipulations of these objects are simplified enormously within our framework, since we already know how 4-d spinors behave in relativistic quantum mechanics. This is one of the main advantages and motivations of the formalism. Important aspects of twistors, such as the physical properties of the massless particle that a twistor is encoding, can be recovered applying the machinery of quantum mechanics to the `translated' 4-d spinor. Furthermore, the conformal geometric algebra enables us to obtain in a simple way the geometrical properties of the twistor. We have also shown, that a twistor can be extended to the 6-d space in the form of a 6-d spinor, with the utility to recover its geometrical properties as observables of a quantum system determined by the 6-d spinor.

It can be argued in this light, that although significant simplifications of the formalism have been achieved, we might be contradicting the initial purpose of twistors, as fundamental objects from which the space of quantum mechanics and general relativity can be constructed, if twistors are taken from the beginning as 4-d spinors. However, in \cite{twistors2} we will show that within our formalism, points in the conformal space can be derived through twistors. Furthermore, we will show that our formalism can bring together multiparticle quantum systems, conformal geometry and wave-functions for spin-0 particles. The relevance of this is that geometric algebra acts as a unifying tool. The goal in the future is to construct wave-functions of different spin within geometric algebra in a unified way, in order to pursue supersymmetry.

\section*{Acknowledgements}

Elsa Arcaute would like to acknowledge the financial support of CONACyT.


\begin{thebibliography}{10}

\bibitem{abl82}
R.~Ablamowicz, Z.~Oziewicz, and J.~Rzewuski.
\newblock {C}lifford algebra approach to twistors.
\newblock {\em J. Math. Phys.}, 23(2):231, 1982.

\bibitem{twistors2}
E.~Arcaute, A.N. Lasenby, and C.J.L. Doran.
\newblock Events and cosmological spaces through twistors in the geometric
  ({C}lifford) algebra formalism.
\newblock Submitted to {\em J. Phys. A-Math Gen}, math-ph/0604048, 2006.

\bibitem{ati64}
M.F. Atiyah, R.~Bott, and A.~Shapiro.
\newblock {C}lifford modules.
\newblock {\em Topology}, 3(Supp. 1):3, 1964.

\bibitem{Bed_etal'05}
J.~Bedford, A.~Brandhuber, B.~Spence, and G.~Travaglini.
\newblock A twistor approach to one-loop amplitudes in {N}=1 supersymmetric
  {Y}ang-{M}ills theory.
\newblock {\em Nucl. Phys. B}, 706(1-2):100--126, 2005.
\newblock hep-th/0410280.

\bibitem{Bena_etal'05}
I.~Bena, Z.~Bern, and D.A. Kosower.
\newblock Twistor-space recursive formulation of gauge-theory amplitudes.
\newblock {\em Phys. Rev. D}, 71(4), 2005.
\newblock hep-th/0406133.

\bibitem{Berkovits'04}
N.~Berkovits.
\newblock Alternative string theory in twistor space for {N}=4
  super-{Y}ang-{M}ills theory.
\newblock {\em Phys. Rev. Lett.}, 93(1), 2004.
\newblock hep-th/0402045.

\bibitem{Berkovits-Witten'04}
N.~Berkovits and E.~Witten.
\newblock Conformal supergravity in twistor-string theory.
\newblock {\em J. High Energy Phys.}, 8(9), 2004.
\newblock hep-th/0406051.

\bibitem{Bern_etal'05}
Z.~Bern, V.~{Del Luca}, L.J. Dixon, and D.A. Kosower.
\newblock All non-maximally-helicity-violating one-loop seven-gluon amplitudes
  in {N}=4 super-{Y}ang-{M}ills theory.
\newblock {\em Phys. Rev. D}, 71(4), 2005.
\newblock hep-th/0410224.

\bibitem{Bidder_etal'05}
S.J. Bidder, N.E.J. Bjerrum-Bohr, L.J. Dixon, and D.C. Dunbar.
\newblock N=1 supersymmetric one-loop amplitudes and the holomorphic anomaly of
  unitary cuts.
\newblock {\em Phys. Lett. B}, 606(1-2):189--201, 2005.
\newblock hep-th/0410296.

\bibitem{Brannan}
D.A. Brannan, M.F. Espleen, and J.J. Gray.
\newblock {\em Geometry}.
\newblock Cambridge University Press, 1999.

\bibitem{Buchbinder}
I.L. Buchbinder and S.M. Kuzenko.
\newblock {\em Ideas and Methods of Supersymmetry and Supergravity, or A Walk
  Through Superspace}.
\newblock IOP, revised edition, 1998.

\bibitem{C-S-W'04_3}
F.~Cachazo, P.~Svrcek, and E.~Witten.
\newblock Gauge theory amplitudes in twistor space and holomorphic anomaly.
\newblock {\em J. High Energy Phys.}, 10(77), 2004.
\newblock hep-th/0409245.

\bibitem{C-S-W'04}
F.~Cachazo, P.~Svrcek, and E.~Witten.
\newblock {MHV} vertices and tree amplitudes in gauge theory.
\newblock {\em J. High Energy Phys.}, 9(6), 2004.
\newblock hep-th/0403047.

\bibitem{C-S-W'04_2}
F.~Cachazo, P.~Svrcek, and E.~Witten.
\newblock Twistor space structure of one-loop amplitudes in gauge theory.
\newblock {\em J. High Energy Phys.}, 10(74), 2004.
\newblock hep-th/0406177.

\bibitem{Chris}
C.J.L. Doran.
\newblock {\em Geometric Algebra and its Application to Mathematical Physics}.
\newblock PhD thesis, University of Cambridge, 1994.

\bibitem{book}
C.J.L Doran and A.N. Lasenby.
\newblock {\em Geometric Algebra for Physicists}.
\newblock Cambridge University Press, 2003.

\bibitem{2}
C.J.L. Doran, A.N. Lasenby, and S.F. Gull.
\newblock States and operators in the spacetime algebra.
\newblock {\em Found. Phys.}, 23(9):1239, 1993.

\bibitem{STA&e}
C.J.L Doran, A.N. Lasenby, S.F. Gull, S.S. Somaroo, and A.D. Challinor.
\newblock Spacetime algebra and electron physics.
\newblock {\em Adv. Imag. \& Elect. Phys.}, 95:271--386, 1996.
\newblock quant-ph/0509178.

\bibitem{FK}
M.~Francis and A.~Kosowsky.
\newblock The construction of spinors in geometric algebra.
\newblock {\em Annals Phys.}, 317:383--409, 2005.
\newblock math-ph/0403040.

\bibitem{hes-sta}
D.~Hestenes.
\newblock {\em Space--Time Algebra}.
\newblock {Gordon and Breach, New York}, 1966.

\bibitem{hes67}
D.~Hestenes.
\newblock Real spinor fields.
\newblock {\em J. Math. Phys.}, 8(4):798, 1967.

\bibitem{bigH}
D.~Hestenes.
\newblock {\em New Foundations for Classical Mechanics}.
\newblock D. Reidel Publishing Company, Dordrecht, Holland, 1986.

\bibitem{hes90}
D.~Hestenes.
\newblock The zitterbewegung interpretation of quantum mechanics.
\newblock {\em Found. Phys.}, 20(10):1213, 1990.

\bibitem{hes-gc}
D.~Hestenes and G.~Sobczyk.
\newblock {\em {C}lifford Algebra to Geometric Calculus}.
\newblock Reidel, Dordrecht, 1984.

\bibitem{anth_geom}
A.N. Lasenby.
\newblock Recent applications of conformal geometric algebra.
\newblock In H.~Li, P.~Olver, and G.~Sommer, editors, {\em Computer Algebra and
  Geometric Algebra with Applications. 6th International Workshop, IWMM 2004,
  Shanghai, China, May, 2004, and International Workshop, GIAE 2004, Xian,
  China, 2004, Revised Selected Papers}, page 298. Springer-Verlag, 2005.

\bibitem{CG_univ}
A.N. Lasenby.
\newblock Conformal geometry and the universe.
\newblock \textit{Phil. Trans. R. Soc. Lond. A}, to appear, 2006.

\bibitem{proc}
A.N. Lasenby, C.J.L. Doran, and E.~Arcaute.
\newblock Applications of geometric algebra in electromagnetism, quantum theory
  and gravity.
\newblock In R~Ab{\l}amowicz, editor, {\em Clifford Algebras, Applications to
  Mathematics, Physics, and Engineering}, pages 467--489. Birkh{\"a}use, 2004.

\bibitem{2-spinors...}
A.N. Lasenby, C.J.L. Doran, and S.F. Gull.
\newblock 2-spinors, twistors and supersymmetry in the spacetime algebra.
\newblock In Z.~Oziewicz, B.~Jancewicz, and A.~Borowiec, editors, {\em Spinors,
  Twistors, {C}lifford Algebras and Quantum Deformations}, page 233. Kluwer
  Academic, Dordrecht, 1993.

\bibitem{Penrose_alg}
R.~Penrose.
\newblock Twistor algebra.
\newblock {\em J. Math. Phys.}, 8:345--366, 1967.

\bibitem{pen69}
R.~Penrose.
\newblock Solutions of the zero-rest-mass equations.
\newblock {\em J. Math. Phys.}, 10(1):38--39, 1969.

\bibitem{pen77}
R.~Penrose.
\newblock The twistor programme.
\newblock {\em Repts. Math. Phys.}, 12(1):65--76, 1977.

\bibitem{pen87}
R.~Penrose.
\newblock On the origins of twistor theory.
\newblock In W.~Rindler and A.~Trautman, editors, {\em Gravitation and
  Geometry: a volume in honour of Ivor Robinson}. Bibliopolis, Naples, 1987.

\bibitem{TP}
R.~Penrose.
\newblock The central programme of twistor theory.
\newblock {\em Chaos, Solitons \& Fractals}, 10(2-3):581--611, 1999.

\bibitem{pen99}
R.~Penrose.
\newblock Twistor theory and the {E}instein vacuum.
\newblock {\em Class. Quantum Grav.}, 16:A113--A130, 1999.

\bibitem{bigPenrose}
R.~Penrose.
\newblock {\em The Road to Reality: A Complete Guide to the Laws of the
  Universe}.
\newblock Jonathan Cape, London, 2004.

\bibitem{pen72}
R.~Penrose and M.A. MacCallum.
\newblock Twistor theory: An approach to the quantisation of fields and
  space-time.
\newblock {\em Phys. Repts.}, 6C:241--315, 1972.

\bibitem{PRI}
R.~Penrose and W.~Rindler.
\newblock {\em Spinors and space-time, Volume I: two-spinor calculus and
  relativistic fields}.
\newblock Cambridge University Press, 1984.

\bibitem{PRII}
R.~Penrose and W.~Rindler.
\newblock {\em Spinors and space-time, Volume II: spinor and twistor methods in
  space-time geometry}.
\newblock Cambridge University Press, 1986.

\bibitem{riesz}
M.~Riesz.
\newblock {\em Clifford Numbers and Spinors}.
\newblock The Institute for Fluid Dynamics and Applied Mathematics, Lecture
  Series No. \textbf{38}, University of Maryland, 1958. Reprinted as fascimile,
  eds. E.F. Bolinder and P. Lounesto. Kluwer Academic Publishers, 1993.

\bibitem{R-V'04}
R.~Roiban and A.~Volovich.
\newblock All conjugate-maximally-helicity-violating amplitudes from
  topological open string theory in twistor space.
\newblock {\em Phys. Rev. Lett.}, 93(13), 2004.
\newblock hep-th/0402121.

\bibitem{Witten'04}
E.~Witten.
\newblock Perturbative gauge theory as a string theory in twistor space.
\newblock {\em Comm. Math. Phys.}, 252(1-3):189--258, 2004.
\newblock hep-th/0312171.

\end{thebibliography}

\end{document}